\documentclass[sigconf, nonacm]{acmart}
\usepackage{booktabs}
\usepackage{graphicx}
\usepackage{subcaption}
\usepackage{xspace}
\usepackage{amsmath,amsfonts,soul}
\sethlcolor{yellow!30}
\usepackage{enumitem}
\usepackage[table]{xcolor}
\usepackage{tcolorbox}
\usepackage{amsmath}
\usepackage{pgfplots}
\pgfplotsset{compat=1.18}
\usepackage{multirow}
\usepackage{tikz}
\usetikzlibrary{positioning,fit}
\usetikzlibrary{arrows.meta}

\newcommand{\stitle}[1]{\vspace{1mm}\noindent{\bf #1.}\xspace}

\newif\ifmohsenfirst
\mohsenfirsttrue

\newcommand{\rvdb}{{\tt rVecDB}\xspace}
\newcommand{\rvdbs}{{\tt rVecDB}s\xspace}
\newcommand{\sse}{{\tt SSE}\xspace}
\newcommand{\sses}{{\tt SSE}s\xspace}

\newcommand{\Reals}{\mathbb{R}}
\newcommand{\simfn}{\mathsf{sim}}

\definecolor{sqlcolor}{RGB}{0,90,90}

\newcommand{\insql}[1]{%
{\color{sqlcolor}\texttt{\footnotesize\bfseries #1}\xspace}
}

\newcommand{\sql}[1]{%
\vspace{1mm}
\begin{flushleft}
\hspace{-2mm} \textcolor{sqlcolor}{\texttt{\footnotesize\bfseries #1}}
\end{flushleft}
\vspace{1mm}
}
\newtcolorbox{examplebox}{
  colback=blue!10,
  colframe=blue!20,
  arc=2mm, 
  fonttitle=\bfseries,
  boxrule=0mm,
  boxsep=1mm,
  left=0mm,
  right=0mm,
  top=0mm,
  bottom=0mm
}
\definecolor{slategray}{RGB}{112, 128, 144}
\definecolor{mutedteal}{RGB}{96, 130, 133}
\definecolor{deepindigo}{RGB}{75, 0, 130}
\definecolor{olivegreen}{RGB}{85, 107, 47}
\definecolor{dustyrose}{RGB}{133, 96, 111}
\definecolor{softblue}{RGB}{70, 130, 180}          
\definecolor{forestgreen}{RGB}{34, 139, 34}        
\definecolor{mutedpink}{RGB}{219, 112, 147}        
\definecolor{burntred}{RGB}{165, 42, 42}           
\definecolor{darkyellow}{RGB}{184, 134, 11}        
\definecolor{warmtaupe}{RGB}{139, 119, 101}        
\definecolor{dustylavender}{RGB}{150, 123, 182}    
\definecolor{steelblue}{RGB}{70, 130, 180}         

\newcommand\vldbdoi{xx.xxxxx/xxxx}
\newcommand\vldbpages{xx -- yy}
\newcommand\vldbvolume{}
\newcommand\vldbissue{}
\newcommand\vldbyear{2027}
\newcommand\vldbauthors{\authors}
\newcommand\vldbtitle{\shorttitle} 
\newcommand\vldbavailabilityurl{https://github.com/UIC-InDeXLab/rVecDB}
\newcommand\vldbpagestyle{empty} 

\begin{document}
\title{Toward Relational-Vector Databases with Schema-Structured Embeddings}


\author{Abolfazl Asudeh}
\orcid{0000-0002-5251-6186}
\affiliation{%
  \institution{University of Illinois Chicago}
}
\email{asudeh@uic.edu}

\author{Mohsen Dehghankar}
\orcid{0009-0006-1687-8012}
\affiliation{%
  \institution{University of Illinois Chicago}
}
\email{mdehgh2@uic.edu}

\author{Mahdi Erfanian}
\orcid{0009-0008-2526-0470}
\affiliation{%
  \institution{University of Illinois Chicago}
}
\email{merfan2@uic.edu}

\begin{abstract}
Vector databases use embeddings for similarity search while relying on traditional attributes for relational query processing. This separation limits their applicability to domains where the semantic attributes required for database operations are not explicitly available.
In this paper, we introduce Relational-Vector Databases (\rvdbs), a database model in which the relational structure is encoded within the embeddings. The enabling representation is the Schema-Structured Embedding (\sse), where different coordinate blocks are geometrically aligned with different semantic facets, and nested prefixes within a block represent different levels of semantic abstraction. Unlike materialized categorical attributes, \sses represent semantic facets as continuous geometric spaces, allowing semantic comparisons beyond a predefined set of labels.
We formulate a range query model over \sses and use it to define relational database operations, including semantic filtering, similarity ranking, and joins. We further formulate grouping and aggregation over the learned semantic spaces. We present two complementary approaches for learning \sses and address schema compatibility across relations through sequential alignment of shared semantic join keys.
Our experiments on four real-world datasets demonstrate the effectiveness of \sses in capturing facet-specific and hierarchical semantics while preserving general similarity search quality. Compared to materialized semantic labels, \sses significantly improves filtering accuracy, achieving a query accuracy comparable to separate facet embeddings while reducing vector storage by a large factor. \sse's block structure further enables efficient query processing, achieving a major speedup over masked embeddings for a composite query on one million tuples.

This paper takes the first step toward the development of \rvdbs, opening up a range of research directions in indexing, query optimization, and physical database design.
The authors hope this work lays the foundation for a new generation of vector databases where relational query processing and similarity search operate directly over learned representations.
\end{abstract}

\maketitle

\pagestyle{\vldbpagestyle}
\begingroup\small\noindent\raggedright\textbf{PVLDB Reference Format:}\\
\vldbauthors. \vldbtitle. PVLDB, \vldbvolume(\vldbissue): \vldbpages, \vldbyear.\\
\href{https://doi.org/\vldbdoi}{doi:\vldbdoi}
\endgroup
\begingroup
\renewcommand\thefootnote{}\footnote{\noindent
This work is licensed under the Creative Commons BY-NC-ND 4.0 International License. Visit \url{https://creativecommons.org/licenses/by-nc-nd/4.0/} to view a copy of this license. For any use beyond those covered by this license, obtain permission by emailing \href{mailto:info@vldb.org}{info@vldb.org}. Copyright is held by the owner/author(s). Publication rights licensed to the VLDB Endowment. \\
\raggedright Proceedings of the VLDB Endowment, Vol. \vldbvolume, No. \vldbissue\ %
ISSN 2150-8097. \\
\href{https://doi.org/\vldbdoi}{doi:\vldbdoi} \\
}\addtocounter{footnote}{-1}\endgroup

\ifdefempty{\vldbavailabilityurl}{}{
\vspace{.3cm}
\begingroup\small\noindent\raggedright\textbf{PVLDB Artifact Availability:}\\
The source code, data, and/or other artifacts have been made available at \url{\vldbavailabilityurl}.
\endgroup
}

\section{Introduction}\label{sec:intro}


Vector databases have become a core infrastructure for modern AI applications. They store high-dimensional embeddings and support approximate nearest neighbor (ANN) search over images, text, multimodal data, and other complex objects. Despite this rapid adoption, the underlying data model remains fundamentally split. Conventional relational attributes, such as time and location, support filtering and structured query processing, while the vector representations (aka embeddings) support similarity search. A constrained ANN query therefore typically has the form
\insql{
    SELECT * FROM  T 
    WHERE {filter}(A,$\tau$) 
    ORDER BY {sim}(E,q)
    LIMIT k.
}
The predicates are evaluated over traditional attributes $A$, while the similarity search (ranking) is evaluated over an embedding $E$~\cite{lin2025survey,pan2024survey,li2025attribute}.

Current vector databases can retrieve objects similar to a query, but they rely on predefined, {explicitly available} traditional attributes to support relational database operations such as filtering, joins, grouping, aggregation, and roll-up/drill-down~\cite{wang2023efficient, zuo2024serf, patel2024acorn, gollapudi2023filtered}.
This separation is increasingly inadequate because, across many domains, such as ecological or medical data, where the metadata required for filtering and other database operations is often incomplete, expensive to curate, or entirely absent~\cite{crandall2023importance, iancu2026assessing, roche2015public}. 
Some other approaches use separate embedding spaces to represent different aspects of an entity, rather than a single general-purpose embedding that supports relational queries over these aspects~\cite{kong2022multi, sun2024multi, milvusmultivector, pgvector}.

In all these cases, embeddings remain opaque: although they may encode rich semantic information, they lack explicit structure, making it difficult to isolate specific concepts or attributes.
In addition, embeddings are difficult to interpret. Although a dense embedding may capture rich information about real-world objects, its individual dimensions generally lack clear, human-understandable meanings. This limits their usefulness in applications where interpretability is important~\cite{wang2025ldir, tehenan2025semantic, senel2021imparting, bhalla2024interpreting}.

\begin{figure*}[!hbt]
\centering
\resizebox{0.9\textwidth}{!}{
\begin{tabular}{c|c||cccc}
\multicolumn{2}{c|}{\textbf{Query}}&\multicolumn{4}{c}{}\\
{\bf similarity} ($q$)&{\bf predicate} ($x$)&\multicolumn{4}{c}{\textbf{Results}}\\ \hline
\includegraphics[width=0.14\linewidth]{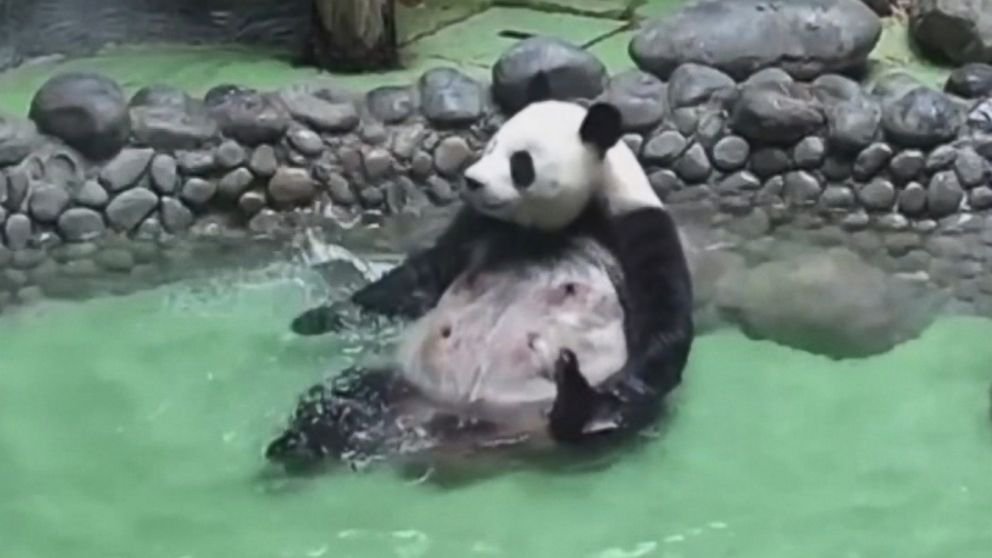} &
\includegraphics[width=0.14\linewidth]{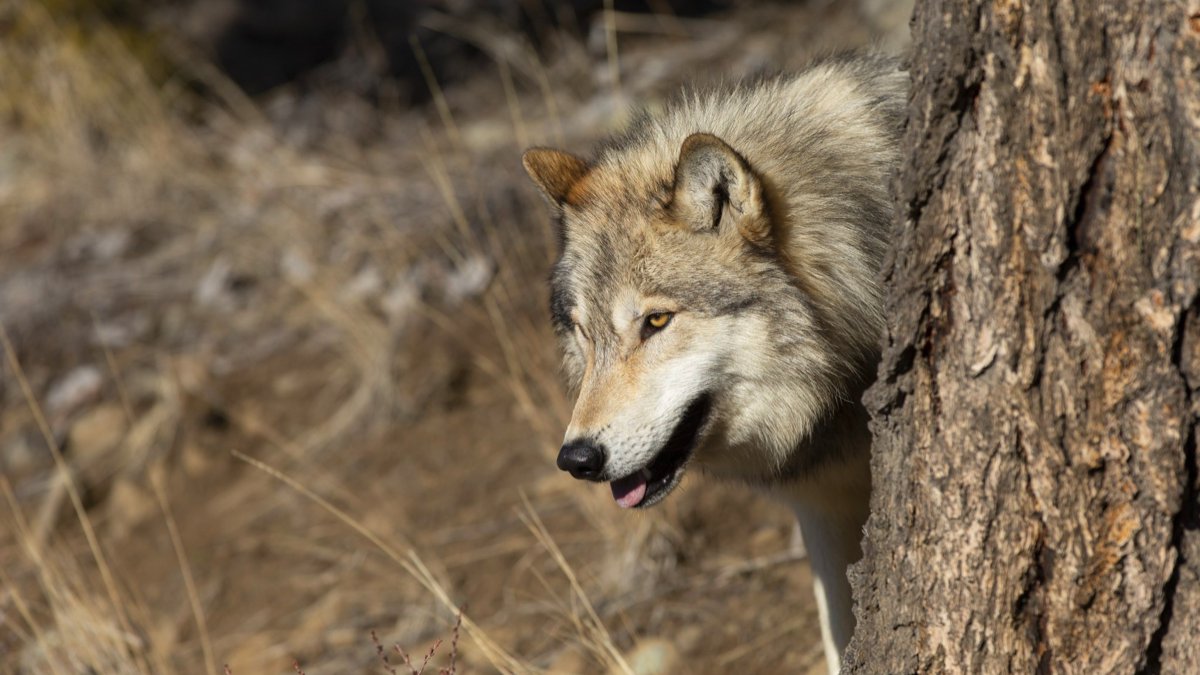} &
\includegraphics[width=0.14\linewidth]{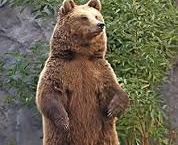} &
\includegraphics[width=0.14\linewidth]{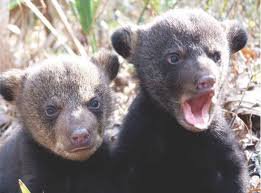} &
\includegraphics[width=0.14\linewidth]{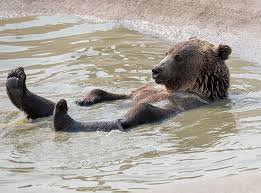} &
\includegraphics[width=0.14\linewidth]{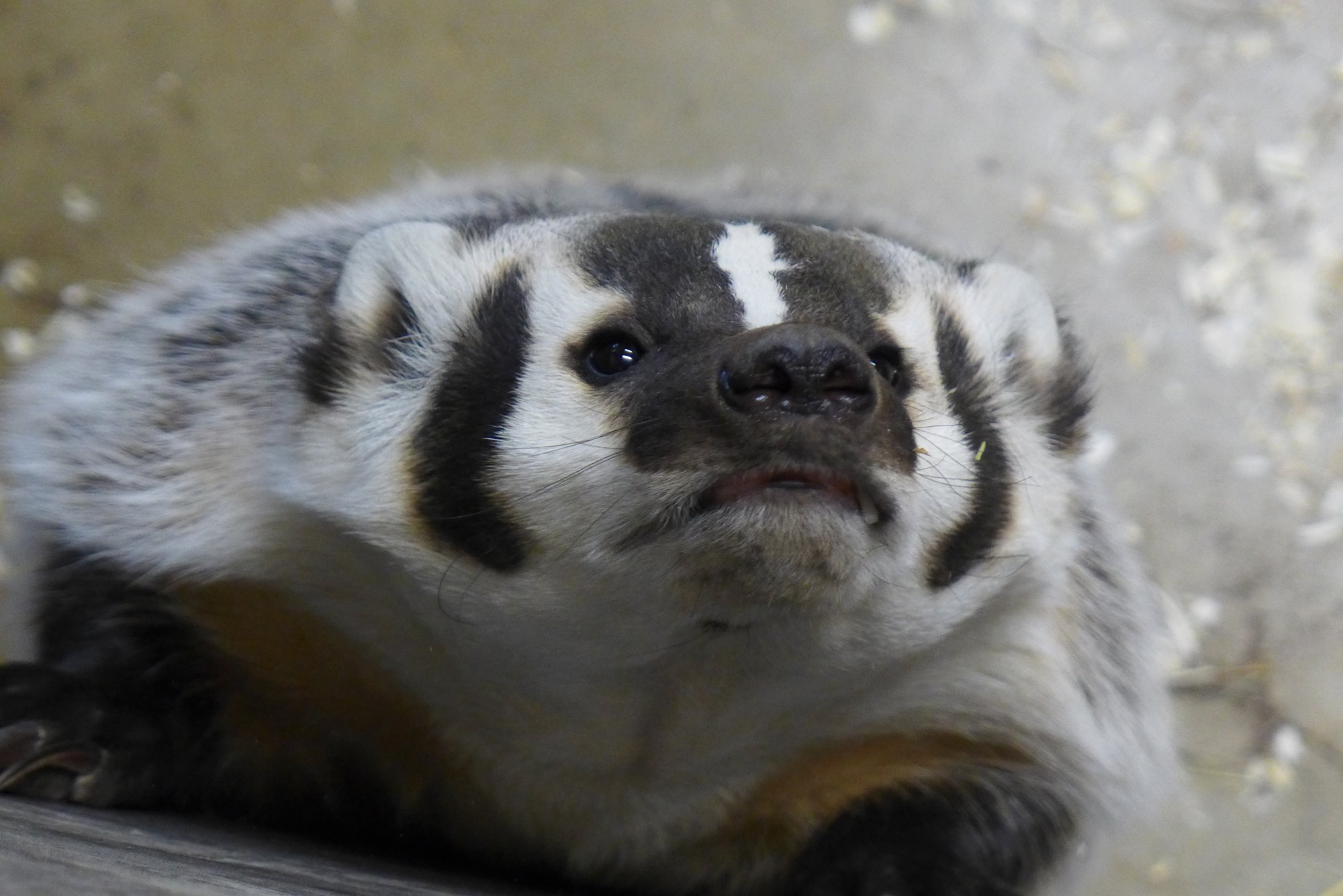} \\
Panda query & Wolf habitat & Bear & Bear & Bear & Badger \\\hline
\includegraphics[width=0.14\linewidth]{animals/panda/4a2ced58ed.jpg} &
\includegraphics[width=0.14\linewidth]{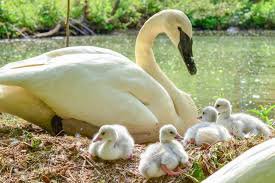} &
\includegraphics[width=0.14\linewidth]{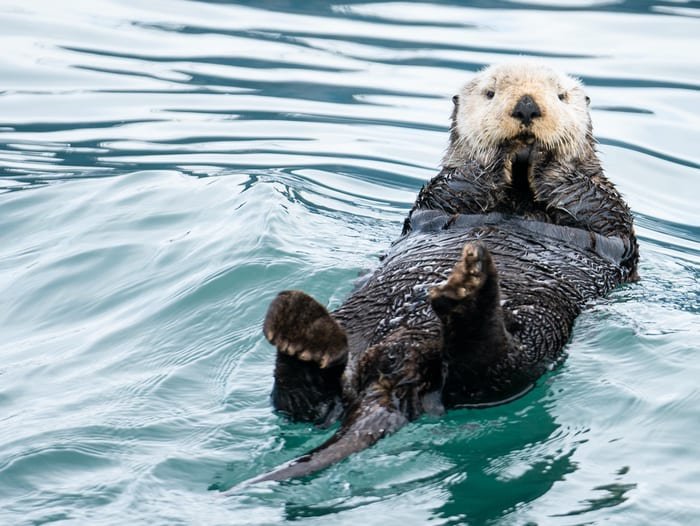} &
\includegraphics[width=0.14\linewidth]{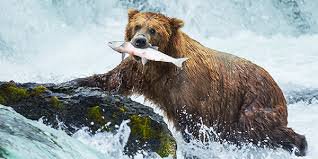} &
\includegraphics[width=0.14\linewidth]{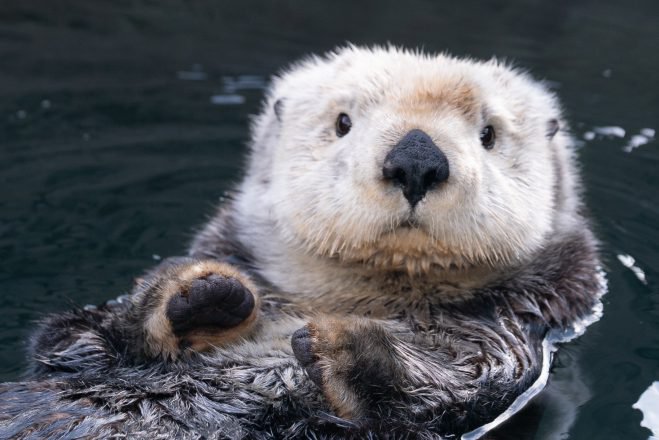} &
\includegraphics[width=0.14\linewidth]{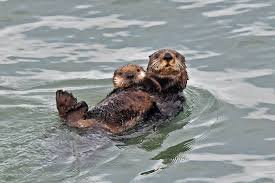} \\
Panda query & Swan habitat & Otter & Bear & Otter & Otter
\end{tabular}
}
\vspace{-3mm} 
\caption{
Retrieved results for Example~\ref{ex-1}.
The query asks for the animal images similar to the query $q$ ({\tt panda}) at the {\tt family} level, 
among the ones in the {\tt habitat} of the predicate $x$.
}
\label{fig:sse-panda}
\vspace{-3mm}
\end{figure*}

In this paper, we propose {\em Relational-Vector Databases} (\rvdbs), a new database model in which {\em embeddings expose a semantic schema structure} that can be used directly in relational query processing. The foundation of \rvdbs is the {\em Schema-Structured Embedding} (\sse). An \sse partitions the embedding into coordinate blocks, each assigned to a semantic facet.
For facets with hierarchical semantics, nested prefixes of a block represent different levels of abstraction. The schema specifies the block structures, while their geometric mapping is learned from data. Unlike conventional vector databases, where relational attributes are stored separately from embeddings, an \rvdb makes the learned semantic structure within the embedding available to database operators.
Let us introduce Example~\ref{ex-1} on a biodiversity dataset containing animal images\footnote{The dataset may optionally contain limited traditional attributes such as id, time, and the information of the photographer.} as the running example across the paper. 

\begin{examplebox}
\begin{example}\label{ex-1}
    On a dataset of biodiversity images, a user may wish to \hl{\tt\small [find $k$ animals with habitats similar to that of reference image (e.g., a swan), ranked by their family-level taxonomic similarity to a panda]}.
    Using \sse (see Section~\ref{sec:sse}), which organizes embeddings into separate blocks for semantic facets such as taxonomy and habitat, and uses nested prefixes within the taxonomy block, this query can be processed by filtering based on similarity in the habitat block and ranking by similarity in the family-level prefix.
    Notably, neither habitat nor taxonomic information is explicitly available as structured attributes in the dataset or the query.
    As we shall explain in Section~\ref{sec:rvdb-model}, \rvdb
    enables formulating the user query as follows, where $x$ is the habitat reference image and $q$ is the ranking reference image:
    \sql{
        \hspace{10mm}SELECT * FROM Animals AS t\\
        \hspace{10mm}WHERE $E_{\mathrm{habitat}}(t)
        \approx_{\tau_h} E_{\mathrm{habitat}}(x)$\\
        \hspace{10mm}ORDER BY $\simfn(E_{\mathrm{tax}}^{(2)}(t),
                         E_{\mathrm{tax}}^{(2)}(q))$
        DESC LIMIT 4
    }
    Figure~\ref{fig:sse-panda} shows our experiment results of this query over a dataset of animal images represented by \sses (see Section~\ref{sec:exp} for more details).
    Changing the predicate tuple while keeping the ranking reference fixed changes the feasible semantic region: a wolf-habitat predicate returns predominantly bear-like animals, while a swan-habitat predicate shifts the results toward otters.
    Interestingly, the returned images often share visual characteristics with the reference panda image.
\end{example}
\end{examplebox}

Indeed, an alternative approach is to train classifiers that predict semantic attributes and materialize their outputs in traditional attributes. Such an approach is useful when the required attributes have a known, fixed vocabulary. However, materialized labels do not preserve the continuous relationships among facet values. \sses instead represent each facet as {\em a continuous geometric space}, allowing predicates to be specified using concept prototypes or example tuples. For instance, a habitat predicate can be defined using the habitat representation of an arbitrary image rather than selecting from a predefined list of habitat categories. This also allows semantic similarity and compatibility to be evaluated at different levels of granularity. Of course \sses do not eliminate the need for supervision, but that they preserve learned semantic relationships as a queryable soft schema in the structure of the vectors rather than discrete traditional labels.

The \rvdb model raises several research questions. First, how can embeddings be learned so that designated blocks reliably capture their intended facets, hierarchical prefixes represent the corresponding abstraction levels, and the full embedding preserves retrieval quality? Second, how should vector indices support constrained ANN when both the filtering predicate and the ranking objective are geometric, potentially over different blocks or prefixes? Third, how can relational operators such as joins, grouping, and aggregation be defined and efficiently evaluated over continuous, learned attributes? These questions connect representation learning with relational database concepts such as database modeling, access methods, and query processing.

This paper takes {\em the first step} by introducing \sse and the \rvdb data and query models.
Specifically, after reviewing the related work in Section~\ref{sec:related}, we first formulate schema-structured embeddings (\sses) in Section~\ref{sec:sse}.
Next, in Section~\ref{sec:rvdb-model}, we define our relational-vector database (\rvdb) model and formally describe the range query model over \sses, which forms the basis for defining database operators.
Our data and query models are agnostic to the techniques used for training \sses. Still, in Section~\ref{sec:learning}, we formulate our learning objective and propose two complementary learning approaches based on the training data at hand.
Joining multiple relations based on semantic blocks requires schema compatibility across those relations. We address this issue in Section~\ref{sec:join-alignment} through sequential alignment of shared semantic join keys.
Next, in Section~\ref{sec:exp}, we provide experimental evidence to validate our approach and evaluate its performance in various settings.
We then provide a set of illustrative queries on our \rvdb model in Section~\ref{sec:queries} to demonstrate its expressiveness and the composition of relational operators over different semantic facets and abstraction levels.
Indeed, the full-stack development of an \rvdb requires addressing a wide range of technical and systems problems, including the
design of access methods that exploit the structure of \sses, selectivity and cardinality estimation over learned semantic spaces, and query planning and optimization.
We briefly discuss some of these research directions in our concluding remarks in Section~\ref{sec:conclusion}.
We hope this paper paves the way for a new generation of vector databases in which learned representations are integral to relational data modeling and query processing.

\section{Related Work}
\label{sec:related}

In this section, we review existing methods for answering queries that combine relational query processing with vector similarity predicates, as illustrated in Example~\ref{ex-1}. We categorize these approaches into two groups based on how they represent and query semantic properties: (1) \textbf{explicit-attribute approaches}, which follow the traditional database paradigm and evaluate predicates over structured attributes or labels, and (2) \textbf{continuous-representation approaches}, a more modern line of work that evaluates predicates through similarity in embedding spaces.

\subsection{Using Explicit Attributes}

As discussed in the introduction, existing filtered ANN search methods combine vector similarity search with predicates over explicitly stored attributes, such as category, location, or time~\cite{gollapudi2023filtered,wang2023efficient,zuo2024serf,patel2024acorn,cai2024navigating,wei2020analyticdb,wang2021milvus}. These methods assume that the attributes required for filtering are explicitly available both in the dataset, as stored fields, and in the query, as specified attribute values.

\textbf{Attributes unavailable in the dataset.} When the required attributes are not stored in the dataset, a common alternative is to train classifiers or concept prediction models that extract semantic attributes from the underlying objects and materialize the predictions as structured fields~\cite{koh2020concept,lu2018accelerating,kang2022tasti,patel2025semantic}. This enables traditional database operations over the predicted labels but has several drawbacks: (i)~prediction errors in the extracted attributes propagate to query results; (ii)~classifiers are restricted to a predefined set of attributes and labels, so they cannot support queries over previously unspecified semantic properties; (iii)~extracting and materializing the labels adds computational and storage costs; and, most importantly, (iv)~discrete labels discard the continuous relationships among attribute values. For example, a habitat classifier may label an image as \emph{wetland} or \emph{forest}, but these labels do not capture how similar two habitats are.

\textbf{Attributes unavailable in the query.} A user may not know the relevant attribute value but may have an example of it, such as another object or a concept prototype; in Example~\ref{ex-1}, the filter is defined by another image without any explicit label. Supporting such queries with a classifier would require mapping each query example to a predefined label at query time, which inherits drawbacks (i)--(iv) above and additionally incurs model inference latency on every query.

In contrast to these methods, \sses represent semantic facets as continuous geometric spaces, so a predicate can be specified directly by a concept prototype or an example object, such as the habitat representation of an arbitrary image. Filtering then becomes similarity search within the corresponding facet, with no labels to predict or materialize.

\subsection{Using Continuous Representations}

Another line of work represents different semantic aspects of an object with aspect-specific embeddings. Multi-Aspect Dense Retrieval~\cite{kong2022multi} and MURAL~\cite{sun2024multi} learn aspect-specific representations for properties such as product category, brand, and color, and vector databases such as Milvus and PostgreSQL with pgvector support multiple vector fields per object~\cite{wang2021milvus,milvusmultivector,pgvector}. Conditional Similarity Networks (CSN)~\cite{veit2017conditional} instead learn a single embedding together with one non-negative mask per notion of similarity, trained from triplet comparisons; similarity under a notion is measured on the masked embedding, and the masks are either learned or fixed to disjoint subsets of dimensions. CSN shows that several notions of similarity can share one embedding, but its masks are flat: it does not represent levels of granularity within an aspect, and it does not treat the masked subspaces as attributes for database operations.

These approaches have three limitations. First, apart from CSN, each aspect has its own embedding model and space, which is redundant when aspects are related, for example nested as in animal taxonomy (order, family, genus): each related attribute is learned and stored independently, multiplying storage and computation while ignoring their shared structure. Matryoshka Representation Learning (MRL)~\cite{kusupati2022matryoshka} nests representations within one embedding: prefixes of increasing length are each usable on their own, and all prefixes are trained with the same objective and labels. A shorter MRL prefix is therefore a cheaper, lower-fidelity version of the same semantics, used to trade accuracy for cost. \sses also use nested prefixes, but each prefix is supervised at its own level of a hierarchy, so a shorter prefix encodes coarser semantics (e.g., order instead of genus) rather than a coarser approximation. Other hierarchical embeddings~\cite{vendrov2016order,nickel2017poincare,ganea2018hyperbolic} capture a hierarchy within one space, but neither they nor MRL organize multiple semantic facets into a unified representation that supports facet-specific predicates.

Second, each embedding is still a high-dimensional dense space whose coordinates have no identifiable meaning. Interpretability matters~\cite{higgins2017beta,subramanian2018spine} when users must understand why an object satisfied a predicate or when results must be audited, and this opacity propagates to downstream systems built on embeddings, from retrieval pipelines to larger models~\cite{lewis2020retrieval}.

Third, the relevant aspects must be identified in advance, which is restrictive when attributes or their values are hard to predefine. For example, consider a dataset of books in which attributes such as author, publication year, and publisher are well-defined, but genre, subgenre, themes, narrative style, mood, and literary tropes are open-ended, often hierarchical, overlapping, and evolving as new styles emerge. A book may be science fiction, more specifically dystopian fiction or cyberpunk, while exploring identity, isolation, and artificial intelligence; a complete set of such attributes and values for all books is hard to define.

In contrast, \sses learn a single, general-purpose embedding that organizes semantic facets into separate blocks, with nested prefixes within a block representing levels of granularity. Related attributes share one space, facets and their hierarchical levels can be queried without separate embeddings, and the dimensions gain an interpretable organization with less redundancy.

\subsection{Query Planning}

Query planning for filtered vector search has been studied extensively: existing methods choose among filter-first, vector-search-first, and interleaved strategies based on filter selectivity and index structure~\cite{wei2020analyticdb,wang2021milvus,patel2024acorn,chronis2025filtered}. These techniques are orthogonal to our work and can be applied to \rvdbs. Moreover, the structure of \sses opens opportunities for new, more efficient query plans, which we discuss in the following sections.
\section{Schema-Structured Embeddings}\label{sec:sse}


Let $t$ denote an object and let $\mathcal{F}=\{\phi_1,\ldots,\phi_m\}$ be a set of semantic facets specified for its domain. A Schema-Structured Embedding (\sse) maps $t$ to a vector $E(t)\in\mathbb{R}^d$ of the form
$$
E(t)=[E_1(t),\ldots,E_m(t),E_{\mathrm{res}}(t)],
$$
where each block $E_i(t)\in\mathbb{R}^{d_i}$ is associated with a semantic facet $\phi_i$, and $E_{\mathrm{res}}(t)$ is an optional residual block. The embedding schema specifies the facets and their corresponding coordinate blocks, including their dimensionalities. These assignments are shared across objects represented under the same schema. The geometric structure within each block is learned from data.

\begin{figure}[!tb]
    \centering
    \resizebox{0.41\textwidth}{!}{
        \begin{tikzpicture}[
    block/.style={
        draw,
        rounded corners=2pt,
        minimum height=1.05cm,
        align=center,
        font=\small
    },
    label/.style={font=\small}
]

\def\h{1.0}
\def\y{0}

\node[block, fill=blue!18,   minimum width=2.25cm] (phylo) at (0,0) {$E_{\mathrm{tax}}(t)$};
\node[block, fill=green!18,  minimum width=2.35cm, right=0pt of phylo] (habitat) {$E_{\mathrm{habitat}}(t)$};
\node[block, fill=orange!22, minimum width=1.85cm, right=0pt of habitat] (dot) {$\cdots$};
\node[block, fill=gray!20,   minimum width=1.75cm, right=0pt of dot] (res) {$E_{\mathrm{res}}(t)$};


\node[label, below=0.0cm of phylo]   {phylogeny};
\node[label, below=0.0cm of habitat] {habitat};
\node[label, below=0.0cm of res]     {residual};


\end{tikzpicture}
    }
    \vspace{-3mm}
    \caption{An \sse of an animal $t$, containing facet-specific blocks, each aligned with different semantic aspects, and the residual block capturing information not assigned to a predefined facet.}
    \label{fig:sse1}
    \vspace{-2mm}
\end{figure}
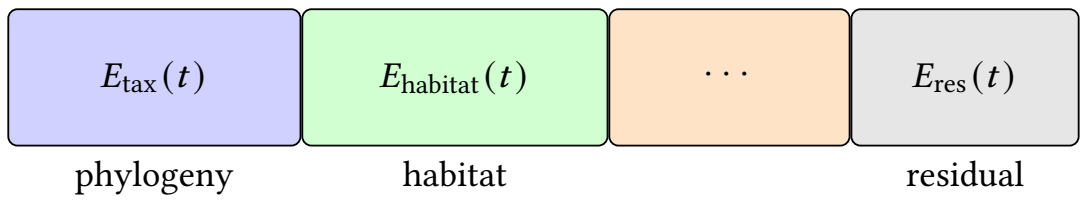

For example, an \sse for an animal image may contain separate blocks for taxonomy, habitat, diet, geography, and morphology,
as illustrated in Figure~\ref{fig:sse1}.
Each block is trained to capture similarity with respect to its designated facet. Two animals with similar habitats should have nearby representations in the habitat block, even if they differ in taxonomy or visual appearance. The taxonomy block, in contrast, should reflect taxonomic relationships.
Different blocks may therefore induce different similarity orderings over the same collection of objects. Note that the facets are not statistically independent, even though each block is trained to represent the relationships relevant to its facet.

The schema specifies which semantic facets are represented, and each block acts as a {\em soft attribute}, whose value is a vector of real numbers rather than a categorical label with a predefined cardinality.
Indeed, training the embeddings may follow predefined labels, but the resulting vector representations are not restricted to those labels.  
That is, two entities with the same labels in the training set, may have similar, but not identical, vector representations on their corresponding blocks.

The residual block provides capacity for retrieval-relevant information not explicitly assigned to the semantic facets. It is optimized jointly with the facet blocks to preserve general similarity-search quality over the full embedding.

\begin{figure}[!tb]
\resizebox{0.41\textwidth}{!}{
    \begin{tikzpicture}[
    >={Latex[length=1.6mm]},
    block/.style={draw, minimum height=0.45cm, minimum width=4.8cm,rounded corners=2pt,
                  align=center, font=\scriptsize},
    arr/.style={<->, thick},
    lab/.style={font=\scriptsize, anchor=west}
]

\node[block, fill=blue!18,] (tax) {$E_{\mathrm{tax}}(t)$};

\draw[arr] (tax.north west) ++(0.00,0.35) -- ++(1.20,0.00);
\node[lab] at ([shift={(1.35,0.35)}]tax.north west)
{$E_{\mathrm{tax}}^{(1)}(t):$ Order};

\draw[arr] (tax.north west) ++(0.00,0.75) -- ++(2.40,0.00);
\node[lab] at ([shift={(2.55,0.75)}]tax.north west)
{$E_{\mathrm{tax}}^{(2)}(t):$ Family};

\draw[arr] (tax.north west) ++(0.00,1.15) -- ([shift={(0.00,1.15)}]tax.north east);
\node[lab] at ([shift={(0.15,1.15)}]tax.north east)
{$E_{\mathrm{tax}}^{(3)}(t):$ Genus};

\end{tikzpicture}
}
\caption{A hierarchical facet block for animal taxonomy.}
\label{fig:sse2}
\end{figure}
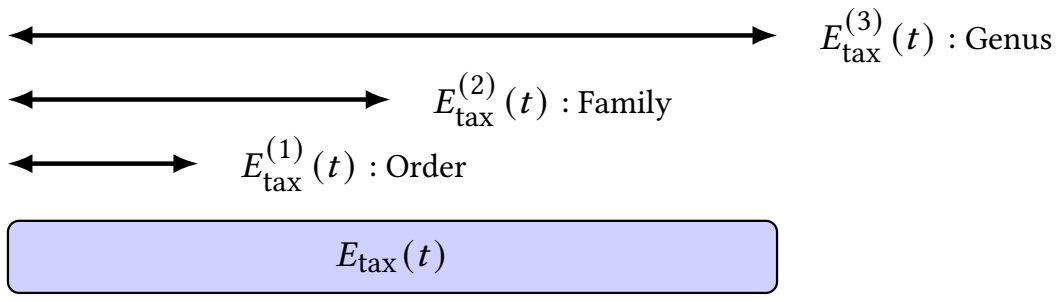

\subsection{Hierarchical Facets}
Some semantic facets have hierarchical structure, with different levels corresponding to different degrees of semantic specificity. For a hierarchical facet $\phi_i$, an \sse assigns {\em nested prefixes} of its block $E_i$ to successive levels of the hierarchy. Let $L_i$ denote the number of levels, with prefix lengths $d_{i,1}<\cdots<d_{i,L_i}=d_i$. The representation of an object $t$ at level $\ell$ is defined as
$$
E_i^{(\ell)}(t)=E_i(t)[1:d_{i,\ell}].
$$
Shorter prefixes represent coarser semantic distinctions, while longer prefixes capture increasingly fine distinctions. The assignment of prefixes to abstraction levels is part of the embedding schema.
Prefix length specifies the semantic abstraction at which objects are compared. Hence, the same facet block can support comparisons at several abstraction levels without storing a separate categorical attribute for each level. 
For example, 
as illustrated in Figure~\ref{fig:sse2},
the taxonomy block of an animal \sse may use dimension prefixes $[1..8]$, $[1..32]$, and $[1..64]$ to represent Order, Family, and Genus, respectively. At the Order level, animals belonging to the same order should have similar representations, even when they belong to different families. Longer prefixes introduce finer distinctions, allowing similarity to be evaluated at the Family or Genus level. 

\subsection{Semantic Comparisons}

Continuous facet representations allow semantic comparisons to be specified through reference vectors rather than only through
predefined labels. A reference may be a learned prototype for a named concept,
a representation constructed from several examples, or the representation of an example object.
In Example~\ref{ex-1},
a habitat comparison may use the habitat block of the \sse of a swan image, i.e., $E_{\mathrm{habitat}}(t_{\mathrm{swan}})$, as its reference.
The comparison then concerns similarity to the habitat represented by that particular example, without requiring the user to assign it a categorical habitat label.

A similarity threshold controls how restrictive a comparison is. Let $E_B(t)$ denote the representation of $t$ in a selected block or hierarchical prefix, and let $c$ be a reference vector in the same space. The condition
$\simfn(E_B(t),c)\geq\tau$ defines a neighborhood around $c$ according to the chosen similarity function. For a fixed space and reference, increasing $\tau$ restricts the matching values, while decreasing it admits a broader range. The same stored representations can therefore support strict or relaxed semantic matching without changing the schema or assigning new labels to the objects.

Deferring the learning of \sses to Section~\ref{sec:learning}, we next introduce our model of the relational-vector databases.

\section{Relational-Vector Database Model}\label{sec:rvdb-model}
We define a Relational-Vector Database (\rvdb) as a set of relations, where each relation $T=\{t_i\}_{i=1}^n$ is a collection of tuples. Each tuple $t\in T$ is represented using a schema-structured embedding (\sse) 
$E(t)\in \Reals^d$, where $E(t)$ is structured into $m<d$ blocks as $E(t) = \left[E_1(t), \cdots, E_m(t), E_{\mathrm{res}}(t)\right]$. Each block $E_\phi(t) \in \Reals^{d_\phi}$ is geometrically aligned with a domain concept or a semantic attribute\footnote{A relation may also contain traditional attributes, such as time, location, or user-provided metadata.}. 
A facet may be non-hierarchical, such as habitat or conservation status, or hierarchical, such as taxonomy, geography, topic, or organizational structure.

\subsection{Query Model}

{\bf Range queries} are central to \rvdb's query model, enabling the database operations on top of \sses.
Formally, let $\simfn:\Reals^{d^*}\rightarrow \mathbb \Reals^+$ be a similarity function that takes two vector representations in
$\Reals^{d^*}$ as the input and measures their similarity, with larger values indicating higher similarities.
Without loss of generality, we assume the similarity values are normalized in range $[0,1]$, such that $\simfn\left(x, x\right)=1,~\forall x\in\Reals^{d^*}$.
When $d^*=d$, the similarity is measured over the entire \sse, while $d^*=d_B< d$ measures the similarity over the projected dimensions of a block (or a hierarchical prefix) $B$.

For a reference vector $c\in \Reals^{d_B}$ in the selected space and thresholds
$0\leq\bot\leq\top\leq 1$, define the {\em semantic range}
$$
\mathcal{R}(T;E_B,c,\bot,\top)
=
\left\{
t\in T:
\bot\leq\simfn(E_B(t),c)\leq\top
\right\}.
$$

For the common case where $\top=1$, we simplify the notation to $\mathcal{R}(T;E_B,c,\bot)=
\left\{
t\in T:
\bot\leq\simfn(E_B(t),c)
\right\}$.

The reference $c$ may be the representation of an example tuple, averaged over several examples, or a learned concept.

Having formulated the range queries, we next show how it enables the database operations on \sses.

\stitle{Semantic predicates}
A semantic predicate is specified as a range query
$$
\phi(E_\phi,c_\phi,\bot,\top)
\equiv
\mathcal{R}(T;E_\phi,c_\phi,\bot,\top).
$$
That is, the tuples that fall inside the range $\mathcal{R}$ satisfy the predicate $\phi$.
For instance, let us consider Example~\ref{eq:obj} with the swan image (2nd row of Figure~\ref{fig:sse-panda}) as the reference image.
The predicate \insql{$E_{\text{Habitat}}\approx_\tau E_{\text{Habitat}}(x)$} is modeled as a range query over the embedding block $E_\phi=E_\text{habitat}$, around the {\tt\small habitat} block of the {\tt\small swan} image, i.e., $c_\phi=E_{\text{Habitat}}(t_\text{swan})$, and a similarity threshold $\tau$. Formally
$$E_{\text{Habitat}}\approx_\tau E_{\text{Habitat}}(x)
\equiv\mathcal{R}(T;E_{\text{Habitat}},E_{\text{Habitat}}(t_\text{swan}),\tau).$$
Similarly, a predicate over a hierarchical block selects a prefix
$E_\phi^{(\ell)}$ and a reference vector in that prefix space.

A conjunction $\Phi=\phi_1\wedge\cdots\wedge\phi_s$ defines
the feasible set
$$
\begin{aligned}
T_\Phi
&=\{t\in T:\phi_j(t)=\mathrm{true},~\forall j\in[s]\}\\
&=\bigcap_{j=1}^{s}
\mathcal{R}
\left(T;E_{\phi_j},c_{\phi_j},\bot_j,\top_j\right).
\end{aligned}
$$
Each predicate may use a different block, prefix, reference, and threshold interval. Disjunction corresponds to union of ranges, while negation is specified using a range query with an upper bound. 
For example, the predicate \insql{$E_{\text{Habitat}}\not \approx E_{\text{Habitat}}(x)$} for the swan image in Example~\ref{ex-1} can be formulated using a similarity upper bound $\tau$ as
$$E_{\text{Habitat}}\not\approx E_{\text{Habitat}}(x)
\equiv\mathcal{R}(T;E_{\text{Habitat}},E_{\text{Habitat}}(t_\text{swan}),0,\tau).$$

Conventional predicates can be combined
with these semantic conditions.

\stitle{Similarity ranking}
A constrained similarity query ranks the feasible tuples
in a selected space $E_\rho$ relative to a query vector $q$.
Let $\tau_k$ be the $k$th largest ranking
score among feasible tuples.
Then the constrained similarity can be formulated as a range query as:
\begin{align*}
    \text{\insql{
    ORDER BY SIM q ON E}} _\rho \text{ \insql{LIMIT k}}
    &=
    \operatorname{Top}_k
    \left(T_\Phi;\, \simfn(E_\rho(t),q)\right)\\
    &\equiv
    \mathcal{R}(T_\phi;E_\rho, q, \tau_k).
\end{align*}
The ranking space $E_\rho$ may coincide with a predicate space,
overlap it, or use different blocks. 
In Example~\ref{ex-1},
the ranking constraint \insql{
{ORDER BY SIM} $E_\text{tax}^{(2)}(q)$ ON $E_\text{tax}^{(2)}$
        {LIMIT } 4
}
returns the top-$4$ tuples based on their similarity with the {\tt panda} image ($q$)
on the {\tt family} level (level 2) of the {\tt taxonomy} block, while the filtering predicate is on the {\tt habitat} block.

Before moving to the next section, we would like to underscore that these definitions specify the logical result, not an
execution order. An implementation may evaluate predicates first, explore the ranking order first, or process the corresponding ranges jointly.

\subsection{From Query Model to Database Operations}\label{sec:querymodel:extended}

\stitle{Semantic Joins}
A semantic {\tt VecJOIN} compares tuples through compatible facet
spaces.
Let $R$ and $S$ be relations with compatible representations
for a selected block or prefix $E_B$. 
A semantic join is defined as
$$
R\bowtie_{B,\bot}S
=
\left\{
(r,s)\in R\times S:
\simfn(E_B(r),E_B(s))\geq \bot
\right\}.
$$
Hence, a join can be formulated using a set of range queries over $S$, where the reference vectors are obtained from the relation
$R$. That is,
$$
R\bowtie_{B,\bot}S
=
\bigcup_{r\in R}
\Big(
\{r\}\times
\mathcal{R}(S;E_B,E_B(r),\bot)
\Big).
$$

For example, animals and habitat records can be joined
through their habitat blocks. Increasing the threshold $\bot=\tau_j$ requires closer habitat similarity, while
decreasing it admits more matches.

Indeed, a semantic {\tt VecJOIN} requires schema compatibility across the joined relations. In Section~\ref{sec:join-alignment}, we address this issue by introducing shared semantic-join keys, which are sequentially aligned during the training process.

\stitle{Grouping}
Unlike grouping by categorical labels, the groups over \sses are determined by continuous facet values.
A semantic {\tt GROUP BY} partitions a relation $T$ into
$k$ groups based on the representations of its tuples in a
selected block or hierarchical prefix $E_B$. Each group $G_i$
is represented by a center $c_i\in\mathbb{R}^{d_B}$, which need
not correspond to an existing tuple. The grouping objective is
$$
\max_{\{G_i,c_i\}_{i=1}^k}
\sum_{i=1}^{k}\sum_{t\in G_i}
\simfn(E_B(t),c_i),
$$
where the $G_i$ form a partition of $T$ into $k$ nonempty groups. Each tuple is assigned to the center with the highest similarity, with ties resolved deterministically. 
That is,
\begin{align*}
    \text{\insql{
    GROUP BY}}_k ~E_B
    \equiv\Big\{
    \{
        t&\in T:\\
    &\simfn(E_B(t),c_i)
        \geq
        \simfn(E_B(t),c_j),\ \forall j\neq i
    \}
    \Big\}_{i=1}^k.
\end{align*}
with ties resolved deterministically.
Thus, \texttt{GROUP BY} induces a partition of the projected embedding space into similarity-based regions, analogous to Voronoi cells.

A group may be represented by its center or by a set of representative tuples that expose its contents to the user.
For example, consider animals grouped by habitat. To represent a group $G_i$, one can cluster its members using their full embeddings, obtaining $h_i$ nonempty subgroups $H_{i,1},\ldots,H_{i,h_i}$. 
We present each subgroup $H_{i,j}$ as the $\langle t^\mu_{H_{i,j}},|H_{i,j}|\rangle$ pair, where $t^\mu_{H_{i,j}}$ is the medoid tuple of $H_{i,j}$.

\stitle{Aggregation}
Aggregation can operate on conventional numeric attributes
or on similarity scores computed from SSEs. Let $U$ denote
the input relation or a group, $E_B$ a selected block or
prefix, and $q$ a reference vector in the corresponding
space. For an aggregate function
$F\in\{\mathrm{SUM},\mathrm{MIN},\mathrm{MAX},\mathrm{AVG}\}$,
we define
$$
\operatorname{Agg}_F(U;E_B,q)
=
F\Big(
\{\simfn(E_B(t),q):t\in U\}
\Big),
$$
where the argument is a multiset of similarity scores, with one contribution per input tuple occurrence.
Please note that the aggregation space may differ from the grouping space.
In Example~\ref{ex-1}, for instance, the grouping can be done using the habitat block, while the aggregate is computed over the family level of the taxonomy block.

In Section~\ref{sec:queries}, we provide a set of query examples to illustrate the composition of the database operations using \rvdb model.

\section{Learning \sse}\label{sec:learning}

\subsection{Learning Objective}

The embedding schema of an \sse specifies the semantic facets, coordinate blocks, and hierarchical levels, but their geometric alignment must be learned from data. An \sse can be obtained by training an encoder directly or by transforming an existing embedding.
For example, given a pretrained vector representation $z(t)$, a learnable projection $f_\theta$ produces
$E(t)=f_\theta(z(t))$.

In order to train an \sse, one can minimize
a loss function that jointly optimizes retrieval preservation, facet alignment, and hierarchical alignment:
\begin{align} \label{eq:obj}
 \mathcal{L} =
\mathcal{L}_{\mathrm{ret}}
+ \sum_{i=1}^{m}\lambda_i
  \mathcal{L}_{\mathrm{facet},i}
+ \sum_{i\in\mathcal{H}}\sum_{\ell=1}^{L_i}
  \mu_{i,\ell}\mathcal{L}_{\mathrm{prefix},i,\ell}~,
\end{align}
where $\mathcal{H}$ denotes the set of hierarchical facets, and the weights $\lambda_i$ and $\mu_{i,\ell}$ are the tunable coefficients that control the contributions of various objectives.
The retrieval loss $\mathcal{L}_{\mathrm{ret}}$ preserves the similarity structure of a pretrained representation or optimizes general retrieval quality. Each facet loss $\mathcal{L}_{\mathrm{facet},i}$ aligns its block with the designated semantic facet, 
while for hierarchical facets, $\mathcal{L}_{\mathrm{prefix},i,\ell}$ encourages alignment at level $\ell$ using supervision from the corresponding hierarchy level. 

Although not limited to any specific training approach, we discuss two complementary approaches for learning the geometric structure of the designated blocks and prefixes below.

\stitle{Pairwise similarity supervision}
Rather than requiring categorical labels, SSEs can be trained using pairwise semantic similarities. Let $\mathcal{P}_B$ be a set of supervised tuple pairs for a block or prefix $E_B$, and let $S_B(t,u)$ denote the target similarity between tuples $t$ and $u$ with respect to the corresponding semantic facet. We minimize the discrepancy between the learned and target similarities:
$$
\mathcal{L}_{\mathrm{sim}}(E_B,S_B)
=
\frac{1}{|\mathcal{P}_B|}
\sum_{(t,u)\in\mathcal{P}_B}
\Big(
\simfn(E_B(t),E_B(u))-S_B(t,u)
\Big)^2.
$$
The targets may be derived from graded annotations, expert-provided similarities, ontologies, or proxy features. For example, the animal images taken in nearby locations share similar habitats. Similarly, given numerical habitat annotations $a_{\mathrm{habitat}}(t)$ generated by an off-the-shelf machine learning model, the target can be defined as the cosine similarity between the annotation vectors. 

Using pairwise similarities, the objective function in Equation~\ref{eq:obj} is instantiated as $\mathcal{L}_{\mathrm{facet},i}=\mathcal{L}_{\mathrm{sim}}(E_i,S_i)$, and $\mathcal{L}_{\mathrm{prefix},i,\ell}=\mathcal{L}_{\mathrm{sim}}(E_i^{(\ell)},S_{i,\ell})$. The retrieval-preservation loss can be instantiated similarly, using the original embedding similarities as targets:
$$
\mathcal{L}_{\mathrm{ret}}
=
\mathcal{L}_{\mathrm{sim}}(E,S_0),
\qquad
S_0(t,u)=\simfn(z(t),z(u)).
$$
This approach follows the principle of similarity-preserving distillation~\cite{tung2019similarity}, which ensures the learned representation preserves pairwise relationships without reproducing the original embedding coordinates.

\stitle{Contrastive supervision}
When supervision identifies semantically related and unrelated examples, we can instead use block-specific or prefix-specific contrastive learning. For a hierarchical facet $i$ at level $\ell$, let $\mathcal{M}$ be a training minibatch and $P_{i,\ell}(t)$ the examples in $\mathcal{M}\setminus\{t\}$ sharing the same label as $t$ at that level. Let $\mathcal{A}_{i,\ell}$ denote the set of tuples $t$ (known as anchors) for which at least one positive example exists in the minibatch at hierarchy level $\ell$. That is $\mathcal{A}_{i,\ell}=\{t\in\mathcal{M}~|~|P_{i,\ell}(t)|>0\}$. Following supervised contrastive learning~\cite{khosla2020supervised}, we define
$$
\begin{aligned}
\mathcal{L}_{i,\ell}^{\mathrm{con}}
={}&-\frac{1}{|\mathcal{A}_{i,\ell}|}
\sum_{t\in\mathcal{A}_{i,\ell}}
\frac{1}{|P_{i,\ell}(t)|}
\sum_{p\in P_{i,\ell}(t)}
\\[-1mm]
&\log
\frac{
\exp\left(\simfn(E_i^{(\ell)}(t),E_i^{(\ell)}(p))/\eta\right)
}{
\sum_{u\in\mathcal{M}\setminus\{t\}}
\exp\left(\simfn(E_i^{(\ell)}(t),E_i^{(\ell)}(u))/\eta\right)
},
\end{aligned}
$$
where $\eta>0$ is a temperature parameter. The loss brings examples sharing the designated semantic property closer while separating examples with different properties. The positive and negative sets depend on the selected facet and hierarchy level. For instance, two animals belonging to different families within the same order are positives at the Order level but negatives at the Family level. 

Note that both of these approaches support incomplete supervision, since the similarity loss uses only pairs for which target similarities are available, and contrastive learning uses available positive and negative relationships. As a result, they do not need to materialize a complete pairwise similarity matrix.

\subsection{Schema Compatibility Across Relations}
\label{sec:join-alignment}

A semantic {\tt VecJOIN} requires the participating blocks to represent the same facet in a common embedding space.
However, this compatibility is not guaranteed when \sses for different relations are learned independently. Even when two blocks represent the same semantic facet and have identical dimensions, their coordinate systems may differ.
For example, under cosine similarity, an orthogonal rotation of either block preserves all within-relation similarities, but generally changes the similarities between tuples from different relations. Consequently, independently learned blocks may capture their respective facets accurately while remaining unsuitable for cross-relation similarity comparisons.

\begin{figure}[t]
\centering
\begin{tikzpicture}[
  every node/.style={font=\scriptsize},
  block/.style={
    draw,
    minimum height=0.45cm,
    align=center,
    inner sep=1pt,
    outer sep=0pt
  },
  key/.style={block,very thick,fill=black!12,
             minimum width=1.2cm}
]

\node[anchor=east] at (1.30,1.25)
    {{\tt ParkSites} ($S$)};
\node[block,minimum width=1.35cm,anchor=east]
    (sgeo) at (2.70,1.25) {geography};
\node[key,anchor=west]
    (shab) at (sgeo.east) {habitat};
\node[block,minimum width=1.15cm,anchor=west]
    (sveg) at (shab.east) {vegetation};
\node[block,minimum width=1.65cm,anchor=west]
    (sres) at (sveg.east) {residual};

\node[block,minimum width=1.65cm,anchor=east]
    (atax) at (2.70,-0.25) {taxonomy};
\node[anchor=east] at (atax.west)
    {{\tt AnimalObs} ($A$)};
\node[key,anchor=west]
    (ahab) at (atax.east) {habitat};
\node[block,minimum width=1.55cm,anchor=west]
    (amorph) at (ahab.east) {morphology};
\node[block,minimum width=0.85cm,anchor=west]
    (ares) at (amorph.east) {residual};

\draw[<->,thick]
    (shab.south) --
    node[right,fill=white,inner sep=1pt]
    {aligned space}
    (ahab.north);

\end{tikzpicture}
\caption{Illustrating two relations with different \sse schemas and a shared habitat join key, where the habitat block of {\tt ParkSites} defines a fixed reference space.}
\label{fig:shared-join-keys}
\end{figure}
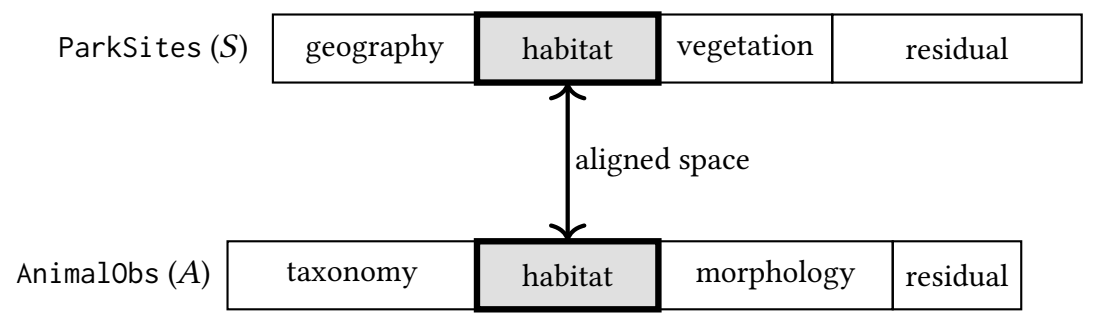

We address this problem through {\em shared semantic join keys}, learned by sequentially aligning the corresponding blocks. For example, consider two relations, {\tt ParkSites} ($S$) and {\tt AnimalObservations} ($A$), as illustrated in Figure~\ref{fig:shared-join-keys}.
The first contains records describing ecological sites in national parks, with facets such as geography, habitat, and vegetation. The second relation contains animal observations, with facets such as taxonomy, habitat, and morphology.
Although the relations have different \sse schemas, their habitat blocks must occupy the same semantic space to support a habitat-based join.
We first learn the \sse of $S$ and use the value of its habitat block $E_{\mathrm{habitat}}^S(s)$ as the ``true label'' when training the \sse of $A$, to align its habitat block with the fixed reference space.

Consider two relations $T$ and $S$, where $T$ refers to $S$ using block $B$ as the join key. 
Let $\mathcal{M}\subseteq T\times S$ denote the known join associations available during training. 
For each $(a,s)\in\mathcal{M}$, the block representation $E_B^S(s)$ serves as the supervision target for $E_B^A(a)$.
Following the existing literature for learning aligned representations from paired observations~\cite{zhai2022lit,girdhar2023imagebind}, we define the alignment loss as
$$
\mathcal{L}_{\mathrm{align}} =
\frac{1}{|\mathcal{M}|}
\sum_{(a,s)\in\mathcal{M}}
\left[
1-\simfn\left(E_B^T(a), E_B^S(s) \right)
\right].
$$
This loss is added to the \sse learning objective, while the reference block remains fixed.
Unlike pairwise similarity supervision, which preserves relative geometric relationships but is invariant to rotations, the alignment loss anchors the learned block to the reference coordinate system.
The remaining blocks are learned according to their respective semantic objectives.

As a result,  the shared key blocks of both relations belong to the same space, allowing the semantic join to be formulated using a set of range queries over $S$, as discussed in Section~\ref{sec:querymodel:extended}.

\subsection{Block Size Allocation}

The dimensionality of each facet block determines its representational capacity and the cost of indexing and querying it. 
Ideally, we would like to allocate the smallest number of dimensions that preserves the quality of semantic comparisons. 
Specifically, let $Q_i(d)$ denote the validation performance of facet $i$ with $d$ dimensions, measured on its intended semantic queries. Given a set of candidate dimensions $\mathcal{D}_i$ and a tolerance $\varepsilon_i$, the block size can be selected as
$$
d_i^* = \min\left\{d\in\mathcal{D}_i: Q_i(d)\geq \max_{d'\in\mathcal{D}_i} Q_i(d')-\varepsilon_i \right\}.
$$
$d_i^*$ selects the smallest dimension whose validation performance is within $\varepsilon_i$ of the best observed performance. For hierarchical facets, the allocation must also account for the performance of the designated prefixes.

Block allocation also supports schema evolution. We consider an append-only model in which the representations of existing facet blocks remain fixed, while new facets are learned using residual capacity or additional dimensions. This preserves existing facet-level predicates and indices without requiring retraining of the original blocks. Changes to the full embedding, however, may affect general similarity search and require updating its index.

\section{Experimental Evaluation}\label{sec:exp}

Having introduced the \rvdb data model and the \sse learning framework, our goal in this section is to evaluate whether a single schema-structured embedding is expressive, accurate, and efficient enough to support relational query processing over unlabeled data. Specifically, we design our evaluation to answer five research questions:
\begin{itemize}[leftmargin=*,topsep=2pt,itemsep=1.5pt]
  \item \textbf{Q1: Representation and hierarchy (Section~\ref{sec:exp:repr}).} Can fixed coordinate blocks and nested prefixes faithfully capture distinct semantic facets and abstraction levels without hurting ordinary full-vector similarity search?
  \item \textbf{Q2: Predicates, ranking, and text queries (Section~\ref{sec:exp:queries}).} How accurately do \sses answer composite range filters, rankings, similarity bands, and natural-language text predicates, especially on unseen classes and graded similarity where explicit labels fail?
  \item \textbf{Q3: Grouping and aggregation (Section~\ref{sec:exp:grouping}).} Can \sses support relational $\texttt{GROUP BY}_k$, representative group summarization (\texttt{REP}), hierarchical roll-up/drill-down, and cross-facet aggregation when no categorical columns exist?
  \item \textbf{Q4: Cross-relation joins (Section~\ref{sec:exp:joins}).} Can \texttt{VecJOIN} link related tuples across tables at a chosen granularity, including tables embedded by \emph{different} encoders and aligned via shared join keys?
  \item \textbf{Q5: System efficiency and robustness (Section~\ref{sec:exp:cost}).} How compact and fast is an \sse in storage, indexing, and query execution compared to alternative architectures, how many dimensions does a block need, and do our findings hold across encoders?
\end{itemize}

\stitle{Summary of observations}
Our experiments answer all five questions in the affirmative and highlight why \sses strike a sweet spot that existing approaches miss. First, compared to pipelines based on materialized classifier labels, zero-shot prompts, plain foundation embeddings, or runtime VLM scoring, \sses are substantially more accurate across every relational operator, particularly on unseen classes, graded compatibility, and zero-shot text references, while running in milliseconds rather than minutes. Second, when multi-vector schemas and masked embeddings (CSN) are upgraded to use our own multi-level \sse losses, \sses match their accuracy with no loss in expressive power while storing up to \textbf{$7.1\times$ less data} than separate per-facet vectors and executing composite queries \textbf{$9\times$ faster} with \textbf{$8.8\times$ smaller facet indices} than masked embeddings.

We first present our experimental setup in Section~\ref{sec:exp:setup}, and then address Q1--Q5 in Sections~\ref{sec:exp:repr}--\ref{sec:exp:cost}.

\subsection{Experimental Setup}\label{sec:exp:setup}

\stitle{Datasets and Workloads}
To test whether \sses generalize across modalities, facet structures, and hierarchy depths, we evaluate on four real-world benchmarks spanning images and text (Table~\ref{tab:datasets}). Every benchmark provides \emph{graded} ground-truth facet similarity 
and holds out unseen leaf classes during training.
\textbf{AwA2}~\cite{xian2019zero} contains $37{,}322$ animal images across $50$ mammal species with continuous human ratings ($0$--$100$) over $85$ semantic properties; we model $14$ habitat properties and $5$ diet properties as continuous facets alongside a three-level taxonomy (Order / Family / Genus), holding out AwA2's standard $10$ test species (e.g., giant panda, seal, leopard) as \emph{unseen} classes.
\textbf{iNaturalist} (\textbf{iNat})~\cite{vanhorn2021benchmarking} ($100{,}000$ images across $10{,}000$ species of animals, plants, and fungi) stresses hierarchical scaling on a deep tree of $273$ orders, $1{,}103$ families, and $4{,}884$ genera, with $20\%$ of species ($2{,}000$) held out as unseen.
\textbf{OpenAlex}~\cite{priem2022openalex} provides a multi-relation scientific benchmark of $99{,}873$ papers (titles and abstracts across all $26$ fields of science) linked to a second table of $5{,}120$ authors across $300$ institutions; each paper follows a three-level topic hierarchy (Domain / Field / Subfield, e.g., \emph{Physical Sciences $>$ Chemical Engineering $>$ Catalysis}), with $20\%$ of leaf topics held out as unseen.
\textbf{MovieLens}~\cite{harper2015movielens} tests non-hierarchical text generality over $13{,}227$ movie plot summaries paired with continuous tag-genome relevance scores~\cite{vig2012tag} grouped into four facets (genre, tone, theme, style).

\begin{table}[!t]
\centering\small
\caption{Benchmark datasets across modalities and domains.}\label{tab:datasets}
\setlength{\tabcolsep}{2.8pt}
\resizebox{\columnwidth}{!}{%
\begin{tabular}{@{}lrrll@{}}
\toprule
Dataset & Tuples & Train / Test & Semantic Facets & Hierarchy Card. \\
\midrule
AwA2 (images) & $37{,}322$ & $24.3\mathrm{k}$ / $13.1\mathrm{k}$ & taxonomy, habitat, diet & $10$ / $27$ / $40$ \\
iNat (images) & $100{,}000$ & $56.0\mathrm{k}$ / $44.0\mathrm{k}$ & taxonomy & $273$ / $1{,}103$ / $4{,}884$ \\
OpenAlex (text) & $99{,}873$ & $56.4\mathrm{k}$ / $43.5\mathrm{k}$ & scientific topic (+ Authors) & $4$ / $26$ / $250$ \\
MovieLens (text) & $13{,}227$ & $9.9\mathrm{k}$ / $3.3\mathrm{k}$ & genre, tone, theme, style & -- \\
\bottomrule
\end{tabular}}
\vspace{-2mm}
\end{table}

\stitle{\sse Architecture and Training}
Each object is encoded once by a frozen foundation model: CLIP ViT-L/14~\cite{radford2021learning} for images (plus EVA02-CLIP-L/14~\cite{sun2023evaclip} in Q5), SPECTER2~\cite{singh2023scirepeval} for scientific papers, and \texttt{bge-base}~\cite{xiao2024cpack} for movie plots and author profiles. A lightweight two-layer MLP projector then maps this vector to a $d{=}768$ \sse (compressed to $d{=}128$ in Q5).
On AwA2, the schema allocates $64$ dimensions to taxonomy (nested prefixes $8 / 32 / 64$ for Order, Family, and Genus), $32$ to habitat, $16$ to diet, and $656$ to the residual block. On the larger hierarchies of iNat and OpenAlex, the taxonomy/topic block uses $128$ dimensions with prefixes $16 / 64 / 128$ and a $640$-dim residual; MovieLens uses four $32$-dim facet blocks and a $640$-dim residual.
We optimize Eq.~\ref{eq:obj}: $\mathcal{L}_{\mathrm{ret}}$ distills the frozen encoder's pairwise similarities; $\mathcal{L}_{\mathrm{facet}}$ aligns each continuous block to ground-truth facet similarities; and $\mathcal{L}_{\mathrm{prefix}}$ combines graded tree-level targets (e.g., at Family: $1.0$ for same family, $0.5$ for same order only, $0$ otherwise) with supervised contrastive loss~\cite{khosla2020supervised} at each prefix's abstraction level.
Projector training takes seconds to a few minutes on a single RTX~5090 GPU. Hyperparameters are selected on a validation split of the training classes, and all results report the mean across three random seeds (standard deviation $\le 0.01$ unless noted).

\stitle{Competing Paradigms}
To ensure a controlled comparison, all baselines share the exact same frozen encoder and training supervision (Table~\ref{tab:baselines}), mirroring the two paradigms from Section~\ref{sec:related}:
\begin{enumerate}[leftmargin=*,topsep=2pt,itemsep=1pt]
  \item \emph{Explicit-attribute and unstructured baselines:}
  \textbf{B1 Labels} represents the standard filtered-ANN database pipeline: classifiers trained on the frozen embedding predict discrete labels for every facet and hierarchy level (e.g., thresholding AwA2 property scores at $50/100$), materializing them as relational columns; facet similarity is label-set overlap and taxonomy similarity is tree-level match ($1, 0.5, 0$).
  \textbf{B2 Prompts} requires no training, scoring each tuple against natural-language text prompts per attribute value (e.g., \emph{``an animal that lives in water''}) and comparing prompt-score profiles.
  \textbf{B3 VLM} evaluates zero-shot multimodal semantic operators~\cite{patel2025semantic} by prompting Qwen2.5-VL-7B~\cite{bai2025qwen25vl} on each reference--candidate image pair ($0$--$10$ rating).
  \textbf{B4 Encoder} uses the unmodified foundation embedding directly for all facets.
  \item \emph{Continuous multi-view architectural baselines:}
  \textbf{B5 Separate Vectors} trains a dedicated MLP and stores a separate embedding per facet and per hierarchy level ($6$ vectors totaling $920$ floats/tuple on AwA2, including the base vector for general search), reflecting multi-vector columns in Milvus~\cite{milvusmultivector} or \texttt{pgvector}~\cite{pgvector}.
  \textbf{B6 CSN}~\cite{veit2017conditional} learns one $768$-dim embedding plus one non-negative $768$-dim mask per facet and level. Crucially, rather than CSN's weaker original triplet loss, we upgrade \emph{both} B5 and B6 with our exact graded and hierarchical \sse losses. Thus, B5 and B6 serve as continuous-learning upper controls that isolate the \emph{architectural and systems cost} of multi-vector fragmentation (B5) and dense overlapping masks (B6) versus a single fixed-block \sse.
\end{enumerate}
For hierarchy analysis (Q1), we also compare two nested-prefix ablations: \textbf{MRL-style}~\cite{kusupati2022matryoshka}, which uses the identical \sse architecture but trains all prefixes ($8,32,64$) on the finest label (Genus/Subfield), and \textbf{PCA Prefix}, which orders encoder coordinates by variance without supervision. In tables, boldface highlights the best result among single-representation database methods (\sse vs.\ B1, B2, B4), while the multi-view controls (B5, B6) are separated to highlight the accuracy--efficiency trade-off.

\begin{table}[!t]
\centering\small
\caption{Competing paradigms. All methods use the same frozen encoder; trained methods use identical supervision.}\label{tab:baselines}
\setlength{\tabcolsep}{2.5pt}
\resizebox{\columnwidth}{!}{%
\begin{tabular}{@{}llll@{}}
\toprule
Method & Paradigm & Stored / Tuple & Facet Evaluation \\
\midrule
\textbf{\sse (ours)} & Schema-structured & $1$ vec ($768$ or $128$d) & Cosine on fixed block / prefix \\
\midrule
B1 Labels & Explicit attributes & $1$ vec + label cols & Predicted label overlap / match \\
B2 Prompts & Zero-shot profile & $1$ vec + score vec & Cosine of prompt-score profiles \\
B3 VLM & LLM/VLM operator & Raw object only & Per-pair VLM inference ($0$--$10$) \\
B4 Encoder & Unstructured vec & $1$ vec ($768$d) & Cosine on full embedding \\
\midrule
B5 Separate & Multi-vector DB & $6$ vecs ($920$d) & Cosine on dedicated facet vector \\
B6 CSN & Learned masks & $1$ vec ($768$d) + masks & Masked $768$d cosine per facet \\
\midrule
MRL-style & Flat nested prefix & $1$ vec ($768$d) & All prefixes trained on leaf class \\
PCA Prefix & Unsupervised prefix & $1$ vec ($768$d) & Top-$k$ principal components \\
\bottomrule
\end{tabular}}
\vspace{-2mm}
\end{table}

\subsection{Representation Quality and Hierarchy (Q1)}\label{sec:exp:repr}

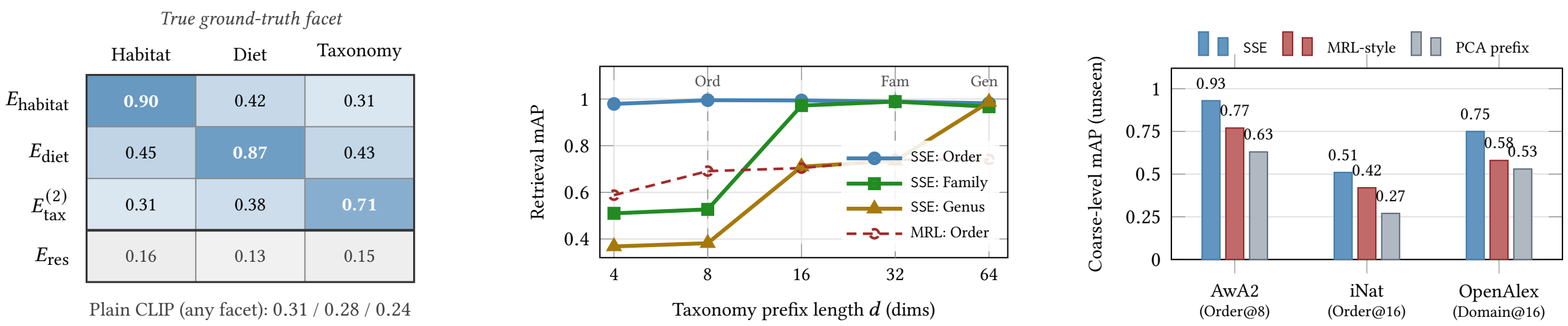
\begin{figure*}[!t]
\centering
\begin{subfigure}[t]{0.32\textwidth}
\centering
\begin{tikzpicture}[font=\scriptsize]
  \def\cw{1.18}
  \def\ch{0.56}
  \fill[steelblue!82] (0*\cw,3*\ch) rectangle (1*\cw,4*\ch);
  \node[white,font=\scriptsize\bfseries] at (0.5*\cw,3.5*\ch) {0.90};
  \fill[steelblue!30] (1*\cw,3*\ch) rectangle (2*\cw,4*\ch);
  \node[black,font=\scriptsize] at (1.5*\cw,3.5*\ch) {0.42};
  \fill[steelblue!20] (2*\cw,3*\ch) rectangle (3*\cw,4*\ch);
  \node[black,font=\scriptsize] at (2.5*\cw,3.5*\ch) {0.31};

  \fill[steelblue!34] (0*\cw,2*\ch) rectangle (1*\cw,3*\ch);
  \node[black,font=\scriptsize] at (0.5*\cw,2.5*\ch) {0.45};
  \fill[steelblue!78] (1*\cw,2*\ch) rectangle (2*\cw,3*\ch);
  \node[white,font=\scriptsize\bfseries] at (1.5*\cw,2.5*\ch) {0.87};
  \fill[steelblue!32] (2*\cw,2*\ch) rectangle (3*\cw,3*\ch);
  \node[black,font=\scriptsize] at (2.5*\cw,2.5*\ch) {0.43};

  \fill[steelblue!20] (0*\cw,1*\ch) rectangle (1*\cw,2*\ch);
  \node[black,font=\scriptsize] at (0.5*\cw,1.5*\ch) {0.31};
  \fill[steelblue!27] (1*\cw,1*\ch) rectangle (2*\cw,2*\ch);
  \node[black,font=\scriptsize] at (1.5*\cw,1.5*\ch) {0.38};
  \fill[steelblue!65] (2*\cw,1*\ch) rectangle (3*\cw,2*\ch);
  \node[white,font=\scriptsize\bfseries] at (2.5*\cw,1.5*\ch) {0.71};

  \fill[gray!14] (0*\cw,0*\ch) rectangle (1*\cw,1*\ch);
  \node[black!75,font=\scriptsize] at (0.5*\cw,0.5*\ch) {0.16};
  \fill[gray!11] (1*\cw,0*\ch) rectangle (2*\cw,1*\ch);
  \node[black!75,font=\scriptsize] at (1.5*\cw,0.5*\ch) {0.13};
  \fill[gray!13] (2*\cw,0*\ch) rectangle (3*\cw,1*\ch);
  \node[black!75,font=\scriptsize] at (2.5*\cw,0.5*\ch) {0.15};

  \draw[black!45,thin] (0,0) grid[xstep=\cw,ystep=\ch] (3*\cw,4*\ch);
  \draw[black!80,thick] (0,1*\ch) -- (3*\cw,1*\ch);
  \draw[black!80,semithick] (0,0) rectangle (3*\cw,4*\ch);

  \node[anchor=east] at (-0.06,3.5*\ch) {$E_{\mathrm{habitat}}$};
  \node[anchor=east] at (-0.06,2.5*\ch) {$E_{\mathrm{diet}}$};
  \node[anchor=east] at (-0.06,1.5*\ch) {$E_{\mathrm{tax}}^{(2)}$};
  \node[anchor=east] at (-0.06,0.5*\ch) {$E_{\mathrm{res}}$};

  \node[anchor=south] at (0.5*\cw,4*\ch+0.04) {Habitat};
  \node[anchor=south] at (1.5*\cw,4*\ch+0.04) {Diet};
  \node[anchor=south] at (2.5*\cw,4*\ch+0.04) {Taxonomy};
  \node[anchor=south,font=\scriptsize\itshape,black!75] at (1.5*\cw,4*\ch+0.38) {True ground-truth facet};

  \node[anchor=north,align=center,font=\scriptsize,black!75] at (1.5*\cw,-0.10)
    {Plain CLIP (any facet): $0.31$ / $0.28$ / $0.24$};
\end{tikzpicture}
\caption{Block--facet correlation (Spearman $\rho$, AwA2).}\label{fig:repr:heatmap}
\end{subfigure}\hfill
\begin{subfigure}[t]{0.33\textwidth}
\centering
\begin{tikzpicture}
\begin{axis}[
    width=1.02\linewidth,
    height=3.65cm,
    xlabel={Taxonomy prefix length $d$ (dims)},
    ylabel={Retrieval mAP},
    xmin=0.85, xmax=5.2,
    xtick={1,2,3,4,5},
    xticklabels={4,8,16,32,64},
    ymin=0.32, ymax=1.14,
    ytick={0.4,0.6,0.8,1.0},
    grid=major,
    grid style={gray!22},
    tick label style={font=\scriptsize},
    label style={font=\scriptsize},
    legend style={font=\tiny, draw=none, fill=white, fill opacity=0.88, text opacity=1, at={(0.98,0.03)}, anchor=south east, inner sep=1pt, row sep=-1pt},
    legend cell align={left}
]
  \draw[dashed,black!35] (axis cs:2,0.32) -- (axis cs:2,1.04);
  \draw[dashed,black!35] (axis cs:4,0.32) -- (axis cs:4,1.04);
  \node[font=\tiny,anchor=south,black!70] at (axis cs:2,1.01) {Ord};
  \node[font=\tiny,anchor=south,black!70] at (axis cs:4,1.01) {Fam};
  \node[font=\tiny,anchor=south,black!70] at (axis cs:4.95,1.01) {Gen};

  \addplot[color=steelblue, very thick, mark=*, mark size=1.5pt] coordinates {
    (1,0.979) (2,0.995) (3,0.994) (4,0.989) (5,0.982)
  };
  \addlegendentry{\sse: Order}

  \addplot[color=forestgreen, very thick, mark=square*, mark size=1.4pt] coordinates {
    (1,0.510) (2,0.527) (3,0.972) (4,0.989) (5,0.967)
  };
  \addlegendentry{\sse: Family}

  \addplot[color=darkyellow!90!black, very thick, mark=triangle*, mark size=1.7pt] coordinates {
    (1,0.368) (2,0.382) (3,0.710) (4,0.735) (5,0.987)
  };
  \addlegendentry{\sse: Genus}

  \addplot[color=burntred, thick, densely dashed, mark=o, mark size=1.4pt] coordinates {
    (1,0.588) (2,0.691) (3,0.704) (4,0.740) (5,0.741)
  };
  \addlegendentry{MRL: Order}
\end{axis}
\end{tikzpicture}
\caption{Prefix zoom staircase (AwA2).}\label{fig:repr:staircase}
\end{subfigure}\hfill
\begin{subfigure}[t]{0.32\textwidth}
\centering
\begin{tikzpicture}
\begin{axis}[
    ybar=1.8pt,
    bar width=5.5pt,
    width=1.02\linewidth,
    height=3.65cm,
    ylabel={Coarse-level mAP (unseen)},
    symbolic x coords={AwA2,iNat,OpenAlex},
    xtick=data,
    xticklabels={AwA2\\[-1pt]{\tiny(Order@8)},iNat\\[-1pt]{\tiny(Order@16)},OpenAlex\\[-1pt]{\tiny(Domain@16)}},
    xticklabel style={align=center,font=\scriptsize},
    ymin=0.0, ymax=1.12,
    ytick={0.0,0.25,0.50,0.75,1.00},
    enlarge x limits=0.24,
    grid=major,
    xmajorgrids=false,
    grid style={gray!22},
    tick label style={font=\scriptsize},
    label style={font=\scriptsize},
    legend columns=3,
    legend style={font=\tiny, draw=none, fill=none, at={(0.5,1.02)}, anchor=south, inner sep=0pt, column sep=2.5pt},
    legend cell align={left},
    nodes near coords,
    every node near coord/.append style={font=\tiny, rotate=0, yshift=0.5pt},
    nodes near coords align={vertical}
]
  \addplot[fill=steelblue!85, draw=steelblue!95!black] coordinates {(AwA2,0.93) (iNat,0.51) (OpenAlex,0.75)};
  \addlegendentry{\sse}

  \addplot[fill=burntred!70, draw=burntred!90!black] coordinates {(AwA2,0.77) (iNat,0.42) (OpenAlex,0.58)};
  \addlegendentry{MRL-style}

  \addplot[fill=slategray!55, draw=slategray!85!black] coordinates {(AwA2,0.63) (iNat,0.27) (OpenAlex,0.53)};
  \addlegendentry{PCA prefix}
\end{axis}
\end{tikzpicture}
\caption{Coarse prefix on unseen classes.}\label{fig:repr:coarse}
\end{subfigure}
\vspace{-2mm}
\caption{Representation quality of \sses. \textbf{(a)}~Each block strongly tracks its designated facet along the diagonal while remaining disentangled from other facets and leaving non-facet signal in $E_{\mathrm{res}}$. \textbf{(b)}~Successive prefixes ($8/32/64$) unlock Order, Family, and Genus step by step while preserving coarser levels; MRL fails to represent Order because all prefixes optimize Genus. \textbf{(c)}~Across image and text hierarchies, level-specific training generalizes coarse prefixes to unseen classes far better than MRL or PCA truncation.}
\label{fig:exp-repr}
\vspace{-2mm}
\end{figure*}

Every \rvdb operator treats coordinate blocks and prefixes as queryable semantic attributes. Before evaluating database queries, we therefore verify three foundational requirements: 1) each block must track its designated facet without leaking other facets, 2) nested prefixes must capture successive levels of a hierarchy, 3) structuring the vector must not hurt ordinary full-vector search.

\stitle{Facet Alignment and Disentanglement}
Over $200\mathrm{k}$ random test pairs on AwA2, we measure how closely each method's facet similarity correlates with ground-truth graded similarity (Table~\ref{tab:facet}). \sse blocks achieve $0.90$ and $0.87$ Spearman correlation on Habitat and Diet, which is nearly \textbf{$3\times$ higher than the unstructured CLIP encoder} ($0.31$ / $0.28$), \textbf{$3.5\times$ higher than zero-shot prompts} ($0.26$ / $0.17$), and $+28$--$29$ points above materialized classifier labels ($0.61$ / $0.59$).
Why do the baselines fail so sharply? Unstructured encoders (B4) entangle all visual features into a single metric: because an otter visually resembles a terrestrial weasel far more than a swan, full-vector similarity groups morphological relatives rather than aquatic animals. Zero-shot text prompts (B2) suffer from CLIP's weak attribute binding on abstract ecological concepts, while binary labels (B1) quantize continuous habitat spectra into coarse pass/fail buckets, discarding degrees of compatibility.

Equally important for relational query composition, Figure~\ref{fig:exp-repr}a shows that \sse coordinate blocks \emph{disentangle} facets cleanly. Each block peaks sharply along the diagonal ($0.90$ Habitat, $0.87$ Diet, $0.71$ Taxonomy) while off-diagonal correlations drop to the natural ecological background between traits ($0.31$--$0.45$; e.g., cetaceans share both taxonomy and marine habitat), and the residual block $E_{\mathrm{res}}$ sheds facet structure ($0.13$--$0.16$) to retain general visual appearance. On MovieLens text, where plot summaries provide a noisier signal and true movie facets naturally overlap, \sse still doubles facet alignment over the plain \texttt{bge-base} embedding across all four facets ($0.25$--$0.35$ vs.\ $0.10$--$0.19$), whereas thresholded binary tag classifiers collapse almost completely ($0.01$--$0.05$). Meanwhile, B5 and B6 match \sse within $0.01$--$0.02$: confining facets to fixed disjoint coordinate slices incurs \emph{zero accuracy penalty} compared to separate networks or full-width masks.

\begin{table}[!t]
\centering\small
\caption{Facet alignment: Spearman $\rho$ with ground-truth graded similarity over $200\mathrm{k}$ test pairs (AwA2).}\label{tab:facet}
\setlength{\tabcolsep}{6pt}
\begin{tabular}{@{}lccc@{}}
\toprule
Method & Habitat & Diet & Taxonomy \\
\midrule
\textbf{\sse (ours)} & \textbf{0.90} & \textbf{0.87} & 0.71 \\
B1 Labels (materialized) & 0.61 & 0.59 & \textbf{0.73} \\
B2 Prompts (zero-shot) & 0.26 & 0.17 & 0.24 \\
B4 Encoder (unstructured) & 0.31 & 0.28 & 0.24 \\
\midrule
B5 Separate vectors (6 vecs) & 0.91 & 0.87 & 0.72 \\
B6 CSN (768d masks) & 0.90 & 0.87 & 0.73 \\
\bottomrule
\end{tabular}
\vspace{-2mm}
\end{table}

\stitle{Hierarchical Zoom Levels via Nested Prefixes}
Table~\ref{tab:prefix} and Figures~\ref{fig:exp-repr}b--\ref{fig:exp-repr}c test whether nested prefixes act as semantic zoom levels across all three hierarchical benchmarks, evaluated on \emph{unseen} species and topics never observed during training.
Figure~\ref{fig:exp-repr}b reveals a striking semantic staircase as more prefix coordinates are read on AwA2: the $8$-dim prefix already saturates Order retrieval ($0.995$ seen / $0.932$ unseen) while leaving different families within an order unseparated ($0.53$); extending to $32$ dimensions unlocks Family ($0.989$) while keeping genera merged ($0.74$); and reading all $64$ dimensions resolves Genus ($0.987$) while \emph{preserving} coarse Order and Family structure inside the longer prefix (pairwise Order AUC $0.95$).

Comparing against MRL-style and PCA prefixes across AwA2, the $4{,}884$-genus iNat tree, and OpenAlex (Table~\ref{tab:prefix}, Figure~\ref{fig:exp-repr}c) exposes a critical representation-learning insight: \emph{coordinate nesting alone does not induce semantic hierarchy}. Because MRL trains every prefix length on leaf-level labels, its $8$-dim prefix is merely a lossy compression of Genus rather than a representation of Order, dropping unseen coarse-level mAP by $16$ points on AwA2 ($0.77$ vs.\ $0.93$), $9$ points on iNat ($0.42$ vs.\ $0.51$), and $17$ points on OpenAlex ($0.58$ vs.\ $0.75$; $0.57$ vs.\ $0.76$ on seen topics). Unsupervised PCA truncation fares even worse ($0.63$ / $0.27$ / $0.53$) because leading variance directions capture background lighting and style rather than domain ontology.
Finally, ablating the prefix objective in Eq.~\ref{eq:obj} shows why combining graded tree similarity with supervised contrastive loss is essential: on deep hierarchies like iNat ($\sim$$5\mathrm{k}$ genera), random pairs rarely share fine taxa, causing pure similarity regression to under-separate fine levels (Genus mAP $0.23$ vs.\ $0.33$), whereas pure contrastive loss pushes sibling families apart and erodes coarse Order geometry inside longer prefixes. Combining both losses achieves the best or tied-best accuracy across every dataset and level.

\begin{table}[!t]
\centering\small
\caption{Hierarchical retrieval mAP on \emph{unseen} species/topics at each level's designated prefix (AwA2: $8/32/64$; iNat \& OpenAlex: $16/64/128$ for Coarse / Mid / Fine).}\label{tab:prefix}
\setlength{\tabcolsep}{2.8pt}
\resizebox{\columnwidth}{!}{%
\begin{tabular}{@{}lccccccccc@{}}
\toprule
 & \multicolumn{3}{c}{AwA2 (O / F / G)} & \multicolumn{3}{c}{iNat (O / F / G)} & \multicolumn{3}{c}{OpenAlex (D / F / S)} \\
\cmidrule(lr){2-4}\cmidrule(lr){5-7}\cmidrule(lr){8-10}
Method & @8 & @32 & @64 & @16 & @64 & @128 & @16 & @64 & @128 \\
\midrule
\textbf{\sse (ours)} & \textbf{0.93} & \textbf{0.98} & \textbf{0.98} & \textbf{0.51} & \textbf{0.36} & \textbf{0.33} & \textbf{0.75} & \textbf{0.46} & \textbf{0.40} \\
MRL-style & 0.77 & 0.95 & 0.97 & 0.42 & 0.35 & 0.32 & 0.58 & 0.44 & 0.39 \\
PCA Prefix & 0.63 & 0.85 & 0.93 & 0.27 & 0.20 & 0.17 & 0.53 & 0.36 & 0.38 \\
\bottomrule
\end{tabular}}
\vspace{-2mm}
\end{table}

\stitle{Preservation of Full-Vector Retrieval}
Imposing block and prefix structure does not compromise ordinary similarity search over the full $768$-dim embedding: same-species / same-topic retrieval mAP matches or slightly exceeds the frozen encoder across AwA2 ($0.82$ vs.\ $0.82$), iNat ($0.12$ vs.\ $0.11$), and OpenAlex ($0.43$ vs.\ $0.41$), retaining $78$--$86\%$ of the encoder's exact top-$10$ neighbors ($74\%$ on MovieLens). Consequently, an \rvdb stores \emph{a single \sse per tuple} for both structured facet operations and unstructured kNN search.

\subsection{Predicates, Ranking, and Text Queries (Q2)}\label{sec:exp:queries}

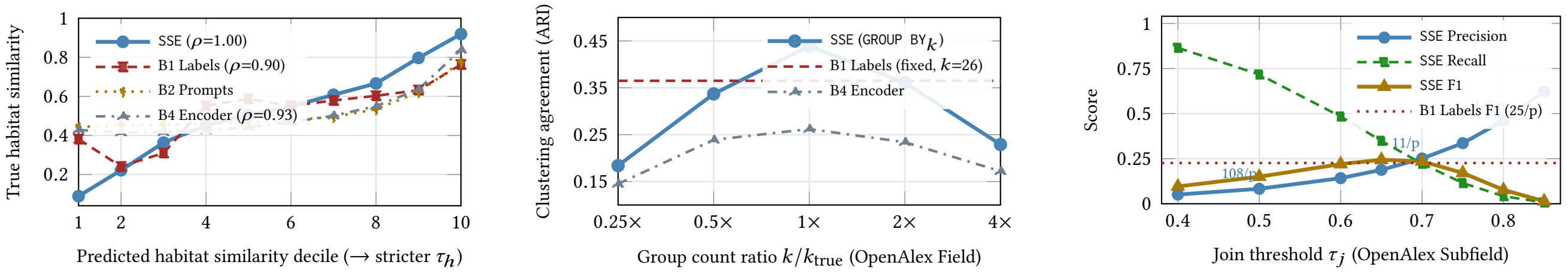
\begin{figure*}[!t]
\centering
\begin{subfigure}[t]{0.32\textwidth}
\centering
\begin{tikzpicture}
\begin{axis}[
    width=1.02\linewidth,
    height=3.65cm,
    xlabel={Predicted habitat similarity decile ($\to$ stricter $\tau_h$)},
    ylabel={True habitat similarity},
    xmin=1, xmax=10,
    xtick={1,2,4,6,8,10},
    ymin=0.04, ymax=1.00,
    ytick={0.2,0.4,0.6,0.8,1.0},
    grid=major,
    grid style={gray!22},
    tick label style={font=\scriptsize},
    label style={font=\scriptsize},
    legend style={font=\tiny, draw=none, fill=white, fill opacity=0.88, text opacity=1, at={(0.03,0.97)}, anchor=north west, inner sep=1pt, row sep=-1pt},
    legend cell align={left}
]
  \addplot[color=steelblue, very thick, mark=*, mark size=1.4pt] coordinates {
    (1,0.089) (2,0.222) (3,0.363) (4,0.452) (5,0.504) (6,0.548) (7,0.608) (8,0.666) (9,0.797) (10,0.919)
  };
  \addlegendentry{\sse ($\rho{=}1.00$)}

  \addplot[color=burntred, thick, densely dashed, mark=square*, mark size=1.3pt] coordinates {
    (1,0.381) (2,0.241) (3,0.310) (4,0.552) (5,0.587) (6,0.551) (7,0.579) (8,0.603) (9,0.632) (10,0.763)
  };
  \addlegendentry{B1 Labels ($\rho{=}0.90$)}

  \addplot[color=darkyellow!90!black, thick, dotted, mark=diamond*, mark size=1.4pt] coordinates {
    (1,0.443) (2,0.453) (3,0.454) (4,0.460) (5,0.467) (6,0.478) (7,0.495) (8,0.534) (9,0.622) (10,0.775)
  };
  \addlegendentry{B2 Prompts}

  \addplot[color=slategray, thick, dashdotted, mark=triangle*, mark size=1.5pt] coordinates {
    (1,0.428) (2,0.415) (3,0.416) (4,0.425) (5,0.443) (6,0.468) (7,0.501) (8,0.550) (9,0.636) (10,0.838)
  };
  \addlegendentry{B4 Encoder ($\rho{=}0.93$)}
\end{axis}
\end{tikzpicture}
\caption{Threshold $\tau_h$ as a continuous dial (AwA2).}\label{fig:ops:dial}
\end{subfigure}\hfill
\begin{subfigure}[t]{0.32\textwidth}
\centering
\begin{tikzpicture}
\begin{axis}[
    width=1.02\linewidth,
    height=3.65cm,
    xlabel={Group count ratio $k / k_{\mathrm{true}}$ (OpenAlex Field)},
    ylabel={Clustering agreement (ARI)},
    xmin=1, xmax=5,
    xtick={1,2,3,4,5},
    xticklabels={$0.25\times$,$0.5\times$,$1\times$,$2\times$,$4\times$},
    ymin=0.10, ymax=0.50,
    ytick={0.15,0.25,0.35,0.45},
    grid=major,
    grid style={gray!22},
    tick label style={font=\scriptsize},
    label style={font=\scriptsize},
    legend style={font=\tiny, draw=none, fill=white, fill opacity=0.88, text opacity=1, at={(0.98,0.97)}, anchor=north east, inner sep=1pt, row sep=-1pt},
    legend cell align={left}
]
  \addplot[color=steelblue, very thick, mark=*, mark size=1.5pt] coordinates {
    (1,0.184) (2,0.337) (3,0.439) (4,0.361) (5,0.229)
  };
  \addlegendentry{\sse ($\texttt{GROUP BY}_k$)}

  \addplot[color=burntred, thick, densely dashed] coordinates {
    (1,0.365) (5,0.365)
  };
  \addlegendentry{B1 Labels (fixed, $k{=}26$)}

  \addplot[color=slategray, thick, dashdotted, mark=triangle*, mark size=1.5pt] coordinates {
    (1,0.145) (2,0.239) (3,0.262) (4,0.234) (5,0.172)
  };
  \addlegendentry{B4 Encoder}
\end{axis}
\end{tikzpicture}
\caption{Impact of $k$ in $\texttt{GROUP BY}_k$ ($k_{\mathrm{true}}{=}26$).}\label{fig:ops:groupk}
\end{subfigure}\hfill
\begin{subfigure}[t]{0.33\textwidth}
\centering
\begin{tikzpicture}
\begin{axis}[
    width=1.02\linewidth,
    height=3.65cm,
    xlabel={Join threshold $\tau_j$ (OpenAlex Subfield)},
    ylabel={Score},
    xmin=0.38, xmax=0.87,
    xtick={0.4,0.5,0.6,0.7,0.8},
    ymin=0.0, ymax=1.04,
    ytick={0.0,0.25,0.5,0.75,1.0},
    grid=major,
    grid style={gray!22},
    tick label style={font=\scriptsize},
    label style={font=\scriptsize},
    legend style={font=\tiny, draw=none, fill=white, fill opacity=0.90, text opacity=1, at={(0.98,0.98)}, anchor=north east, inner sep=1pt, row sep=-1pt},
    legend cell align={left}
]
  \addplot[color=steelblue, very thick, mark=*, mark size=1.3pt] coordinates {
    (0.40,0.051) (0.50,0.083) (0.60,0.142) (0.65,0.188) (0.70,0.251) (0.75,0.336) (0.80,0.461) (0.85,0.622)
  };
  \addlegendentry{\sse Precision}

  \addplot[color=forestgreen, thick, densely dashed, mark=square*, mark size=1.2pt] coordinates {
    (0.40,0.865) (0.50,0.717) (0.60,0.484) (0.65,0.349) (0.70,0.222) (0.75,0.115) (0.80,0.042) (0.85,0.007)
  };
  \addlegendentry{\sse Recall}

  \addplot[color=darkyellow!90!black, very thick, mark=triangle*, mark size=1.5pt] coordinates {
    (0.40,0.096) (0.50,0.150) (0.60,0.220) (0.65,0.244) (0.70,0.235) (0.75,0.171) (0.80,0.077) (0.85,0.014)
  };
  \addlegendentry{\sse F1}

  \addplot[color=burntred, thick, dotted] coordinates {
    (0.38,0.226) (0.87,0.226)
  };
  \addlegendentry{B1 Labels F1 (25/p)}

  \node[font=\tiny,steelblue!90!black,anchor=south east,inner sep=0.5pt] at (axis cs:0.50,0.11) {108/p};
  \node[font=\tiny,steelblue!90!black,anchor=south east,inner sep=0.5pt] at (axis cs:0.70,0.28) {11/p};
  \node[font=\tiny,steelblue!90!black,anchor=south east,inner sep=0.5pt] at (axis cs:0.805,0.49) {1.2/p};
\end{axis}
\end{tikzpicture}
\caption{\texttt{VecJOIN} selectivity vs.\ precision.}\label{fig:ops:join}
\end{subfigure}
\vspace{-2mm}
\caption{Continuous query behavior across \rvdb operators. \textbf{(a)}~Sweeping $\tau_h$ across similarity deciles monotonically increases true habitat compatibility from $0.09$ to $0.92$, whereas predicted labels oscillate and plateau at $0.76$ and the plain encoder remains flat across lower deciles. \textbf{(b)}~$\texttt{GROUP BY}_k$ peaks at the true cardinality and smoothly rolls up ($k<k_{\mathrm{true}}$) or drills down ($k>k_{\mathrm{true}}$), surpassing fixed label buckets. \textbf{(c)}~Varying $\tau_j$ in \texttt{VecJOIN} smoothly trades recall for precision and output size (matches per paper annotated), whereas a label equality join offers only one rigid operating point.}
\label{fig:exp-ops}
\vspace{-2mm}
\end{figure*}

\begin{table*}[!t]
\centering\small
\caption{Relational-vector query and aggregation benchmark on AwA2 ($1{,}000$ composite queries, $300$ aggregation queries). \textbf{Filter (All / Seen / Unseen)}: F1 of habitat range predicate $E_{\mathrm{habitat}}(t)\approx_{\tau_h} E_{\mathrm{habitat}}(x)$ split by whether reference $x$ belongs to a seen or unseen species; \textbf{Rank}: NDCG@10 of Family-prefix ranking relative to $q$; \textbf{Mixing}: feasible-set Jaccard error when references share habitat but differ in taxonomy ($\downarrow$ lower is better); \textbf{Graded}: Spearman $\rho$ among pairs whose binary habitat labels differ; \textbf{Band}: F1 of interval predicate $\bot_h \le \simfn \le \top_h$; \textbf{Agg (AVG / MAX)}: Spearman $\rho$ of per-Order habitat aggregates (Section~\ref{sec:exp:grouping}).}\label{tab:queries}
\setlength{\tabcolsep}{5.2pt}
\begin{tabular}{@{}lccccccccc@{}}
\toprule
 & \multicolumn{3}{c}{Habitat Filter F1 $\uparrow$} & Family & Facet & Graded & Band & \multicolumn{2}{c}{Order Agg.\ ($\rho$) $\uparrow$} \\
\cmidrule(lr){2-4}\cmidrule(lr){9-10}
Method & All & Seen $x$ & Unseen $x$ & Rank $\uparrow$ & Mixing $\downarrow$ & Sim.\ $\uparrow$ & Pred.\ $\uparrow$ & \texttt{AVG} & \texttt{MAX} \\
\midrule
\textbf{\sse (ours)} & \textbf{0.75} & \textbf{0.79} & \textbf{0.72} & \textbf{0.56} & \textbf{0.15} & \textbf{0.89} & \textbf{0.60} & \textbf{0.77} & \textbf{0.85} \\
B1 Labels (materialized) & 0.48 & 0.50 & 0.47 & 0.38 & 0.35 & 0.59 & 0.35 & 0.55 & 0.66 \\
B2 Prompts (zero-shot) & 0.47 & 0.43 & 0.49 & 0.41 & 0.40 & 0.21 & -- & -- & -- \\
B4 Encoder (unstructured) & 0.55 & 0.48 & 0.61 & 0.39 & 0.35 & 0.23 & 0.34 & 0.53 & 0.70 \\
\midrule
B5 Separate vectors (6 vecs) & 0.76 & 0.79 & 0.73 & 0.57 & 0.15 & 0.89 & 0.61 & 0.77 & 0.85 \\
B6 CSN (768d masks) & 0.75 & 0.79 & 0.73 & 0.57 & 0.15 & 0.89 & 0.60 & 0.77 & 0.87 \\
\bottomrule
\end{tabular}
\vspace{-2mm}
\end{table*}

Having verified that individual blocks and prefixes capture their intended semantics, we now ask whether composing them in end-to-end queries, such as filtering on one facet while ranking on another, selecting similarity bands, or specifying predicates in natural language, delivers accurate answers, especially in regimes where explicit labels fail.

\stitle{Composite Predicates, Ranking, and Range Bands}
We run $1{,}000$ composite queries modeled on Example~\ref{ex-1} (and Q1--Q2 in Section~\ref{sec:queries}): select tuples whose habitat block is close to a reference animal $x$, then rank the feasible set by Family-level similarity to an independent query animal $q$ (Table~\ref{tab:queries}; similar gains hold on MovieLens text, where composite F1 reaches $0.35$ vs.\ $0.18$ for labels).
\sse outperforms materialized labels (B1) by \textbf{$+56\%$ relative in Filter F1} ($0.75$ vs.\ $0.48$) and \textbf{$+47\%$ in Ranking NDCG@10} ($0.56$ vs.\ $0.38$), while cutting cross-facet mixing error more than in half ($0.15$ vs.\ $0.35$--$0.40$): when a swan and an otter serve as habitat references, \sse returns consistent aquatic sets despite their distant taxonomies, whereas when a polar bear and a giant panda serve as references, \sse separates arctic from bamboo-forest habitats despite their shared bear family ($0.12$ error vs.\ $0.54$ for B1).
We also evaluate the VLM semantic operator (\textbf{B3}, Qwen2.5-VL-7B) on a $50$-query subset: pairwise VLM scoring is \emph{both substantially less accurate} than \sse (Filter F1 $0.53$ vs.\ $0.75$) and \emph{computationally prohibitive}, processing only $16.3$ image pairs/s on an RTX~5090 and requiring $\sim$$25$~minutes per query over the AwA2 test split compared to milliseconds for \sse.

Table~\ref{tab:queries} and Figure~\ref{fig:exp-ops}a pinpoint three structural reasons why explicit-attribute pipelines fail on semantic queries:
\begin{enumerate}[leftmargin=*,topsep=2pt,itemsep=1pt]
  \item \emph{Open-world unseen classes:} When the reference $x$ belongs to a species never labeled during training (e.g., seal or giant panda), classifiers must force it into seen label buckets ($0.47$ F1), whereas \sse projects visual features directly into continuous habitat geometry, retaining nearly all of its seen-species accuracy ($0.79 \to 0.72$).
  \item \emph{Graded similarity and band predicates:} Among pairs whose binary habitat labels \emph{disagree}, \sse still orders true ecological compatibility with $\rho{=}0.89$ (vs.\ $0.59$ for B1 and $0.23$ for B4). This continuous geometry directly enables interval band predicates (\emph{``moderately similar habitat''}, $\bot_h \le \simfn \le \top_h$), nearly doubling F1 over both labels and the plain encoder ($0.60$ vs.\ $0.35$ / $0.34$).
  \item \emph{Calibrated threshold control:} As shown in Figure~\ref{fig:exp-ops}a, tightening $\tau_h$ across deciles monotonically raises true habitat similarity from $0.09$ to $0.92$ ($\rho{=}1.00$). By contrast, discrete label overlap oscillates non-monotonically across tie-buckets and caps at $0.76$, while the plain encoder stays flat ($\approx\!0.42$) across the bottom five deciles.
\end{enumerate}

\stitle{Zero-Shot Cross-Modal Text Predicates}
Because CLIP aligns images and text in a shared backbone, an \rvdb user need not supply a reference image at all: a predicate can be specified in natural language. Without training on any text, we encode a species prompt (\emph{``a photo of a polar bear''}), apply a standard mean-offset shift to bridge CLIP's modality gap~\cite{liang2022mind}, pass it through the image-trained \sse projector, and query $E_{\mathrm{habitat}}$.
As shown in Table~\ref{tab:textq}, zero-shot text references recover nearly the full accuracy of image references ($0.71$ vs.\ $0.79$ F1 across all $50$ species) and outperform raw CLIP text-to-image search by \textbf{$+22$ F1 points} ($0.71$ vs.\ $0.49$), including on unseen species ($0.60$ vs.\ $0.53$). Raw CLIP text search retrieves animals that \emph{look like} a polar bear; projecting the text vector into $E_{\mathrm{habitat}}$ isolates \emph{where a polar bear lives}, returning whales, seals, and walruses.

\begin{table}[!t]
\centering\small
\caption{Zero-shot cross-modal habitat filtering on AwA2: natural-language text reference vs.\ image reference.}\label{tab:textq}
\setlength{\tabcolsep}{6.5pt}
\begin{tabular}{@{}lccc@{}}
\toprule
Query Modality \& Space & All ($50$) & Seen ($40$) & Unseen ($10$) \\
\midrule
\textbf{\sse}, image reference & \textbf{0.79} & \textbf{0.81} & \textbf{0.73} \\
\textbf{\sse}, text reference (zero-shot) & \textbf{0.71} & \textbf{0.73} & \textbf{0.60} \\
Raw CLIP, image reference & 0.49 & 0.47 & 0.58 \\
Raw CLIP, text reference & 0.49 & 0.48 & 0.53 \\
\bottomrule
\end{tabular}
\vspace{-2mm}
\end{table}

\subsection{Semantic Grouping and Aggregation (Q3)}\label{sec:exp:grouping}

Beyond selecting and ranking individual tuples, a relational query engine must partition unlabeled collections into meaningful groups and compute group-level aggregates. We next test whether $\texttt{GROUP BY}_k$, representative summarization (\texttt{REP}), hierarchical drill-down, and cross-facet aggregation work reliably over continuous \sse blocks.

\begin{figure}[!t]
\centering
\includegraphics[width=\columnwidth]{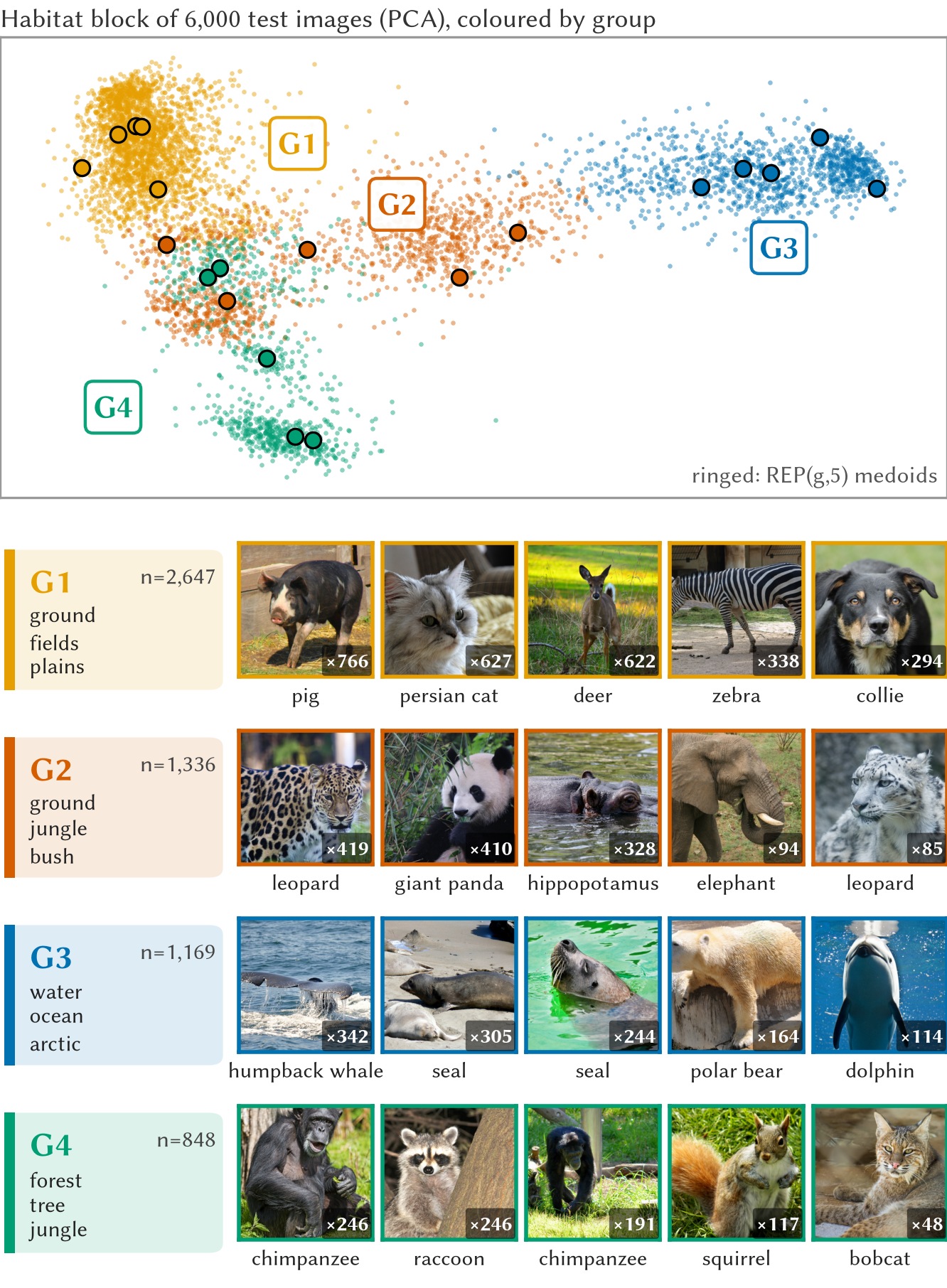}
\vspace{-5mm}
\caption{Execution of \texttt{SELECT REP(g,5), COUNT(*) FROM Animals GROUP BY$_4$ $E_{\mathrm{habitat}}$} on $6{,}000$ AwA2 test images. \textbf{Top:} 2D PCA projection of $E_{\mathrm{habitat}}$ colored by discovered group, with \texttt{REP(g,5)} medoids ringed. \textbf{Bottom:} each group (annotated on the left with size and members' top true properties for reference) summarized by five full-embedding medoids and subgroup counts.}\label{fig:rep}
\vspace{-3mm}
\end{figure}

\stitle{Semantic Grouping and Roll-Up/Drill-Down}
We execute $\texttt{GROUP BY}_k$ via spherical $k$-means on the chosen block or prefix (Section~\ref{sec:querymodel:extended}), and implement hierarchical drill-down by recursively partitioning Order groups into Family and Genus subgroups across AwA2, iNat, and OpenAlex (Table~\ref{tab:grouping}).
On continuous facets such as AwA2 habitat, grouping by predicted binary label combinations shatters animals into $27$ brittle bit-patterns (NMI $0.41$, ARI $0.17$), whereas $\texttt{GROUP BY}_4$ on $E_{\mathrm{habitat}}$ discovers coherent ecological biomes at NMI $0.68$ (ARI $0.72$, vs.\ $0.51$ / $0.43$ for B4). Figure~\ref{fig:rep} visualizes the output of \texttt{SELECT REP(g,5), COUNT(*)} (Q4 in Section~\ref{sec:queries}): without any stored labels, \sse partitions the $6{,}000$ test images into plains and ground, jungle and bush, marine and arctic, and forest habitats, summarizing each group via five full-embedding medoids and their represented tuple counts.

Across taxonomic and scientific hierarchies, \sse prefixes consistently beat the plain encoder at every granularity (e.g., iNat bird Orders $0.50$ vs.\ $0.41$; OpenAlex Fields $0.57$ vs.\ $0.45$). Materialized classifiers (B1) are competitive \emph{only} when every test group is a closed-world class seen during training (AwA2 seen Orders: $0.89$ vs.\ $0.78$). Whenever data contain \emph{unseen} species or topics, which is the realistic regime for exploratory grouping, classifiers cannot invent new group IDs and collapse (predicting only $49$ of $224$ unseen bird genera on iNat), whereas \sse forms accurate groups directly from geometry (\textbf{$0.75$ vs.\ $0.47$ NMI} on iNat unseen genera; \textbf{$0.69$ vs.\ $0.51$} on OpenAlex unseen subfields).
Moreover, Figure~\ref{fig:exp-ops}b shows that varying $k$ provides a smooth granularity knob: agreement peaks at $k_{\mathrm{true}}$ (OpenAlex Field ARI $0.44$ vs.\ $0.37$ for fixed labels and $0.26$ for B4), while smaller $k$ rolls up related fields and larger $k$ drills down into high-purity subtopics (AwA2 habitat NMI: $0.48$, $0.71$, $0.68$, $0.63$, and $0.60$ for $k{=}2, 3, 4, 8, 16$).

\stitle{Cross-Facet Aggregation}
The rightmost columns of Table~\ref{tab:queries} evaluate cross-facet aggregation (Q5 in Section~\ref{sec:queries}): computing per-Order \texttt{AVG} and \texttt{MAX} habitat similarity to $300$ reference animals, where Order membership itself is determined by the $8$-dim Order prefix. Remarkably, even though B1 assigns Order membership as accurately as \sse ($0.94$ vs.\ $0.93$), \sse ranks Order-level aggregates far more accurately (\texttt{AVG} $\rho{=}0.77$ vs.\ $0.55$; \texttt{MAX} $0.85$ vs.\ $0.66$) because continuous habitat scores preserve subtle within-group variation that binary labels erase.

\begin{table}[!t]
\centering\small
\caption{Semantic $\texttt{GROUP BY}_k$ agreement (NMI; $k{=}k_{\mathrm{true}}$) across graded facets, seen hierarchies, and unseen leaf classes.}\label{tab:grouping}
\setlength{\tabcolsep}{2.3pt}
\resizebox{\columnwidth}{!}{%
\begin{tabular}{@{}lcccccc@{}}
\toprule
 & \multicolumn{2}{c}{AwA2} & \multicolumn{2}{c}{iNat (Birds)} & \multicolumn{2}{c}{OpenAlex} \\
\cmidrule(lr){2-3}\cmidrule(lr){4-5}\cmidrule(lr){6-7}
Method & Habitat & Order & Order & Genus$^\dagger$ & Field & Subfield$^\dagger$ \\
\midrule
\textbf{\sse (ours)} & \textbf{0.68} & 0.78 & \textbf{0.50} & \textbf{0.75} & \textbf{0.57} & 0.69 \\
B1 Labels (fixed groups) & 0.41 & \textbf{0.89} & 0.30 & 0.47 & 0.52 & 0.51 \\
B4 Encoder & 0.51 & 0.53 & 0.41 & 0.66 & 0.45 & \textbf{0.70} \\
\midrule
B5 Separate vectors & 0.64 & 0.80 & 0.50 & 0.75 & 0.57 & 0.69 \\
B6 CSN & 0.65 & 0.81 & 0.53 & 0.75 & 0.57 & 0.69 \\
\bottomrule
\end{tabular}}\\[0.5mm]
{\footnotesize $^\dagger$Evaluated exclusively on species/topics held out from training.}
\vspace{-2mm}
\end{table}

\subsection{Cross-Relation Semantic Joins (Q4)}\label{sec:exp:joins}

A relational database must also connect tuples across distinct tables. Here we ask whether \texttt{VecJOIN} can match related entities across relations at a chosen abstraction level, both when the two relations share a common encoder and in the harder, realistic setting where each relation is embedded by a different model and trained independently.

\stitle{Multi-Granularity Joins and Semantic Near-Misses}
On OpenAlex, we join \texttt{Papers} with \texttt{Authors} across Domain ($4$), Field ($26$), and Subfield ($250$) prefixes ($16/64/128$ dims) to answer: \emph{``for each paper, which $10$ authors work closest to its research topic?''} (ground truth: authors whose primary publication area matches the paper's area in OpenAlex's taxonomy).
We first test homogeneous schema alignment, where an author's \sse is the mean \sse of their other papers (excluding the query paper, with no author supervision). As shown in Table~\ref{tab:joins} (left), \sse outperforms predicted-label joins and raw SPECTER2 across every granularity: at the $250$-class Subfield level, \sse reaches P@10 of $0.36$ (\textbf{$40\times$ random} at $0.009$ and \textbf{$1.8\times$ predicted labels} at $0.20$) and maintains the same lead on unseen topics ($0.37$ vs.\ $0.21$).

Strict exact-label scoring penalizes scientifically close matches (e.g., matching an AI paper with a Computer Vision researcher counts the same as matching Manual Therapy). We therefore evaluate partial semantic credit at the Subfield level using two independent domain-relatedness keys: \emph{Text} (cosine similarity of \texttt{bge-large} embeddings of subfield descriptions, calibrated so identical subfields score $1$ and pairs at or below the median score $0$, e.g., AI$\leftrightarrow$Vision $0.32$, Oncology$\leftrightarrow$Cancer Research $0.40$) and \emph{Co-activity} (empirical rate at which authors publish in both subfields). Under both keys, \sse preserves a clear margin ($0.41$ vs.\ $0.32$ for labels on \texttt{Papers}$\bowtie$\texttt{Authors}; $0.62$ vs.\ $0.52$ on \texttt{Papers}$\bowtie$\texttt{Papers}), showing that its non-exact matches land overwhelmingly in adjacent research areas.

\begin{table}[!t]
\centering\small
\caption{\texttt{Papers} $\bowtie$ \texttt{Authors} \texttt{VecJOIN} (P@10) on OpenAlex under (left) shared-encoder aggregation and (right) separately trained \sses across \emph{different} encoders (SPECTER2 $\bowtie$ \texttt{bge-base}) on held-out authors. \emph{Sem.}: Subfield P@10 with text semantic credit.}\label{tab:joins}
\setlength{\tabcolsep}{2.2pt}
\resizebox{\columnwidth}{!}{%
\begin{tabular}{@{}lcccc@{\hspace{6pt}}cccc@{}}
\toprule
 & \multicolumn{4}{c}{Homogeneous Encoder} & \multicolumn{4}{c}{Cross-Encoder ($\mathcal{L}_{\mathrm{align}}$)} \\
\cmidrule(lr){2-5}\cmidrule(lr){6-9}
Method & Dom. & Fld. & Sub. & Sem. & Dom. & Fld. & Sub. & Sem. \\
\midrule
\textbf{\sse (aligned)} & \textbf{0.81} & \textbf{0.53} & \textbf{0.36} & \textbf{0.41} & \textbf{0.81} & \textbf{0.47} & \textbf{0.24} & \textbf{0.35} \\
\sse (no $\mathcal{L}_{\mathrm{align}}$) & -- & -- & -- & -- & 0.20 & 0.04 & 0.01 & 0.08 \\
B1 Labels (equality) & 0.75 & 0.48 & 0.20 & 0.32 & 0.70 & 0.41 & 0.18 & 0.26 \\
B4 Raw encoders & 0.78 & 0.45 & 0.30 & 0.36 & 0.30 & 0.04 & 0.01 & 0.08 \\
Random & 0.27 & 0.04 & 0.01 & 0.07 & 0.28 & 0.04 & 0.01 & 0.07 \\
\bottomrule
\end{tabular}}
\vspace{-2mm}
\end{table}

\stitle{Heterogeneous Encoders and Shared Join Keys}
The decisive test of Section~\ref{sec:join-alignment} occurs when two relations use \emph{different} encoders and train their \sses separately. Keeping the \texttt{Papers} \sse (SPECTER2) fixed as the reference relation $S$, we embed \texttt{Authors} ($T$) from raw publication-title profiles using a distinct model, \texttt{bge-base}, and train the \texttt{Authors} \sse on $70\%$ of authors using authorship links as join associations $\mathcal{M}$ in $\mathcal{L}_{\mathrm{align}}$, evaluating on the held-out $30\%$ unseen authors (Table~\ref{tab:joins}, right).
Two findings stand out. First, \emph{without explicit coordinate alignment, cross-relation joins fail completely}: an \texttt{Authors} \sse trained with the exact same hierarchical losses on its own labels, but omitting $\mathcal{L}_{\mathrm{align}}$, organizes authors well internally, yet its cross-relation join collapses to random chance ($0.04$ Field, $0.01$ Subfield), because pairwise similarity losses are invariant to orthogonal rotations between independently learned spaces. Raw encoders (SPECTER2 vs.\ \texttt{bge-base}) fail identically ($0.04$).
Second, adding $\mathcal{L}_{\mathrm{align}}$ anchors the two schemas into a common geometric frame, boosting held-out join precision by \textbf{$12\times$ at Field} ($0.47$) and \textbf{$24\times$ at Subfield} ($0.24$), surpassing predicted-label equality joins at every level ($0.81/0.47/0.24$ vs.\ $0.70/0.41/0.18$, and $0.26$ vs.\ $0.18$ on unseen topics).

\stitle{Selectivity--Precision Control via $\tau_j$}
Unlike an equality join on predicted labels, which emits a single rigid output ($25$ authors/paper at Subfield, Precision $0.17$, F1 $0.23$), the threshold $\tau_j$ of \texttt{VecJOIN} exposes a continuous selectivity--precision trade-off (Figure~\ref{fig:exp-ops}c). Sweeping $\tau_j$ from $0.50$ to $0.80$ smoothly contracts output cardinality from $108$ to $1.2$ authors/paper while lifting precision from $0.08$ to $0.46$; at $\tau_j{=}0.65$--$0.70$, \sse matches or exceeds the label join's F1 ($0.24$ vs.\ $0.23$) while materializing \textbf{fewer than half as many join pairs} ($11$ vs.\ $25$ per paper). At coarser levels (Field), while top-$k$ \texttt{VecJOIN} clearly beats labels ($0.47$ vs.\ $0.41$), a single uncalibrated global $\tau_j$ achieves lower peak F1 ($0.33$ vs.\ $0.41$) because similarity neighborhoods vary in density across broad disciplines, pointing to per-query threshold calibration as an open query-optimization problem for \rvdbs.

\subsection{System Efficiency and Robustness (Q5)}\label{sec:exp:cost}

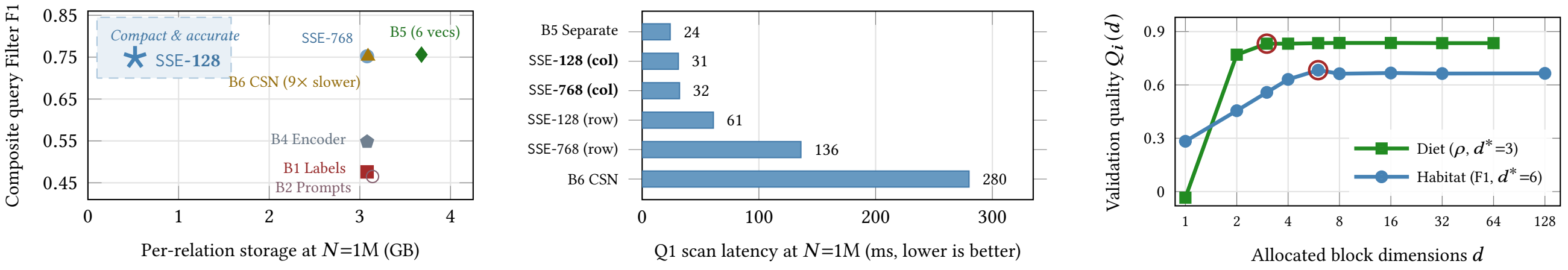
\begin{figure*}[!t]
\centering
\begin{subfigure}[t]{0.32\textwidth}
\centering
\begin{tikzpicture}
\begin{axis}[
    width=1.02\linewidth,
    height=3.65cm,
    xlabel={Per-relation storage at $N{=}1\mathrm{M}$ (GB)},
    ylabel={Composite query Filter F1},
    xmin=0.0, xmax=4.25,
    ymin=0.41, ymax=0.86,
    xtick={0,1,2,3,4},
    ytick={0.45,0.55,0.65,0.75,0.85},
    grid=major,
    grid style={gray!22},
    tick label style={font=\scriptsize},
    label style={font=\scriptsize}
]
  \fill[steelblue!12, draw=steelblue!45, densely dashed] (axis cs:0.10,0.70) rectangle (axis cs:1.58,0.845);
  \node[font=\tiny\itshape,steelblue!90!black,anchor=north west] at (axis cs:0.13,0.84) {Compact \& accurate};

  \addplot[only marks, color=steelblue, mark=star, mark options={scale=1.8, very thick}] coordinates {(0.52,0.749)};
  \node[font=\scriptsize\bfseries,steelblue!90!black,anchor=west] at (axis cs:0.63,0.742) {\sse-128};

  \addplot[only marks, color=steelblue!75, mark=*, mark size=2.0pt] coordinates {(3.08,0.752)};
  \node[font=\tiny,steelblue!85!black,anchor=south east] at (axis cs:3.04,0.760) {\sse-768};

  \addplot[only marks, color=forestgreen!85!black, mark=diamond*, mark size=2.3pt] coordinates {(3.68,0.756)};
  \node[font=\tiny,forestgreen!80!black,anchor=south] at (axis cs:3.72,0.766) {B5 (6 vecs)};

  \addplot[only marks, color=darkyellow!90!black, mark=triangle*, mark size=2.3pt] coordinates {(3.09,0.752)};
  \node[font=\tiny,darkyellow!80!black,anchor=north east] at (axis cs:3.12,0.730) {B6 CSN ($9\times$ slower)};

  \addplot[only marks, color=slategray, mark=pentagon*, mark size=2.1pt] coordinates {(3.08,0.548)};
  \node[font=\tiny,slategray!90!black,anchor=east] at (axis cs:2.96,0.556) {B4 Encoder};

  \addplot[only marks, color=burntred, mark=square*, mark size=1.9pt] coordinates {(3.08,0.476)};
  \node[font=\tiny,burntred!90!black,anchor=east] at (axis cs:2.96,0.486) {B1 Labels};

  \addplot[only marks, color=dustyrose, mark=o, mark size=1.9pt] coordinates {(3.14,0.465)};
  \node[font=\tiny,dustyrose!95!black,anchor=east] at (axis cs:3.02,0.436) {B2 Prompts};
\end{axis}
\end{tikzpicture}
\caption{Accuracy vs.\ storage ($N{=}1\mathrm{M}$).}\label{fig:sys:pareto}
\end{subfigure}\hfill
\begin{subfigure}[t]{0.33\textwidth}
\centering
\begin{tikzpicture}
\begin{axis}[
    xbar,
    bar width=5.0pt,
    width=1.00\linewidth,
    height=3.65cm,
    xlabel={Q1 scan latency at $N{=}1\mathrm{M}$ (ms, lower is better)},
    symbolic y coords={csn,sse768row,sse128row,sse768col,sse128col,b5},
    ytick=data,
    yticklabels={B6 CSN,\sse-768 (row),\sse-128 (row),\textbf{\sse-768 (col)},\textbf{\sse-128 (col)},B5 Separate},
    yticklabel style={font=\tiny},
    xmin=0, xmax=335,
    xtick={0,100,200,300},
    enlarge y limits=0.14,
    grid=major,
    ymajorgrids=false,
    grid style={gray!22},
    tick label style={font=\scriptsize},
    label style={font=\scriptsize},
    nodes near coords,
    every node near coord/.append style={font=\tiny, anchor=west, xshift=1pt}
]
  \addplot[fill=steelblue!82, draw=steelblue!95!black] coordinates {
    (280,csn)
    (136,sse768row)
    (61,sse128row)
    (32,sse768col)
    (31,sse128col)
    (24,b5)
  };
\end{axis}
\end{tikzpicture}
\caption{Physical layout \& scan latency ($N{=}1\mathrm{M}$).}\label{fig:sys:latency}
\end{subfigure}\hfill
\begin{subfigure}[t]{0.32\textwidth}
\centering
\begin{tikzpicture}
\begin{axis}[
    width=1.02\linewidth,
    height=3.65cm,
    xlabel={Allocated block dimensions $d$},
    ylabel={Validation quality $Q_i(d)$},
    xmin=-0.2, xmax=7.3,
    xtick={0,1,2,3,4,5,6,7},
    xticklabels={1,2,4,8,16,32,64,128},
    ymin=-0.08, ymax=0.98,
    ytick={0.0,0.3,0.6,0.9},
    grid=major,
    grid style={gray!22},
    tick label style={font=\tiny},
    label style={font=\scriptsize},
    legend style={font=\tiny, draw=none, fill=white, fill opacity=0.88, text opacity=1, at={(0.98,0.05)}, anchor=south east, inner sep=1pt, row sep=-1pt},
    legend cell align={left}
]
  \addplot[color=forestgreen, very thick, mark=square*, mark size=1.3pt] coordinates {
    (0,-0.034) (1,0.770) (1.585,0.832) (2,0.832) (2.585,0.835) (3,0.836) (4,0.836) (5,0.835) (6,0.835)
  };
  \addlegendentry{Diet ($\rho$, $d^*{=}3$)}

  \addplot[color=steelblue, very thick, mark=*, mark size=1.4pt] coordinates {
    (0,0.283) (1,0.455) (1.585,0.558) (2,0.631) (2.585,0.683) (3,0.663) (4,0.667) (5,0.664) (7,0.665)
  };
  \addlegendentry{Habitat (F1, $d^*{=}6$)}

  \addplot[only marks, color=burntred, mark=o, mark size=2.8pt, thick] coordinates {
    (1.585,0.832) (2.585,0.683)
  };
\end{axis}
\end{tikzpicture}
\caption{Block-size diminishing returns (AwA2).}\label{fig:sys:blocksize}
\end{subfigure}
\vspace{-2mm}
\caption{System efficiency and schema sizing. \textbf{(a)}~\sse-128 uniquely occupies the compact, high-accuracy corner ($7.1\times$ smaller than B5 Separate Vectors; $+27$~F1 points above B1 Labels; B3 VLM requires $\sim$25\,min/query at F1~$0.53$). \textbf{(b)}~Fixed schema offsets allow block-columnar storage, cutting scan latency by $2\times$--$4.2\times$ ($31$--$32$\,ms regardless of residual width) and beating B6 CSN by $9\times$. \textbf{(c)}~Facet quality saturates at $d^*{=}3$--$6$ dimensions (red circles), confirming low intrinsic facet dimensionality.}
\label{fig:exp-sys}
\vspace{-2mm}
\end{figure*}

In Q1--Q4, continuous multi-view baselines (B5 Separate Vectors and B6 CSN) matched \sse in accuracy once upgraded with our training losses. This raises a practical systems question: what does \rvdb gain from organizing everything into a single fixed-block \sse? To answer this, we evaluate physical storage, scan latency, index footprint, block-size selection, and encoder independence.

\stitle{Physical Layout, Storage, and Query Latency at Scale}
First, keeping the $112$ facet dimensions fixed on AwA2 and compressing only the residual block from $656$ to $144$ (\sse-256) or $16$ dimensions (\sse-128) leaves facet and composite query accuracy \emph{unchanged} (Filter F1 $0.752 \to 0.749$, Rank NDCG $0.560 \to 0.562$, Graded $\rho$ $0.889 \to 0.889$, unseen Order@8 mAP $0.932 \to 0.934$) while retaining strong full-vector retrieval (mAP $0.80$ vs.\ $0.82$).

Table~\ref{tab:cost} and Figures~\ref{fig:exp-sys}a--\ref{fig:exp-sys}b benchmark physical storage, exact-scan latency of composite query Q1 (habitat filter at $25\%$ selectivity + Family top-$10$ ranking), and \texttt{faiss} HNSW~\cite{malkov2018efficient} ($M{=}32$) facet indexing over $N{=}1{,}000{,}000$ tuples:
\begin{itemize}[leftmargin=*,topsep=2pt,itemsep=1pt]
  \item \textbf{Advantage over Separate Vectors (B5):} B5 fragments every tuple across $6$ separate vectors ($920$ floats $= 3.68$\,GB), requiring separate columns and model heads for every facet and hierarchy level. \sse-128 consolidates the entire schema and residual into \textbf{one compact vector ($0.52$\,GB, $7.1\times$ smaller)}. Moreover, because \sse block offsets are fixed by the schema, storing blocks in a \emph{block-columnar} physical layout (zero space overhead) lets Q1 scan only the $32$ habitat floats and $32$ Family-prefix floats per passing tuple. This cuts scan latency by \textbf{$2\times$--$4.2\times$} over row-major storage ($61 \to 31$\,ms for \sse-128; $136 \to 32$\,ms for \sse-768, making query speed independent of residual width) and matches B5 within $7$\,ms while building identical $32$-dim HNSW facet indices ($0.38$\,GB, $0.02$\,ms top-$10$ lookup).
  \item \textbf{Advantage over Masked Embeddings (B6 CSN):} Inspecting CSN's learned masks reveals why soft masking is ill-suited for databases: rather than isolating sparse coordinates, each CSN mask activates $365$--$435$ of the $768$ dimensions with $\sim\!50\%$ overlap between any two facets. Because coordinates cannot be sliced, every CSN facet evaluation requires a full $768$-dim masked inner product, making Q1 \textbf{$9\times$ slower} ($280$ vs.\ $31$\,ms), and indexing a single facet requires a $768$-dim HNSW graph (\textbf{$8.8\times$ larger}: $3.33$ vs.\ $0.38$\,GB; $4.7\times$ slower build; $3\times$ slower search). Even pre-materializing CSN's active dimensions per facet balloons relation storage to \textbf{$11.02$\,GB ($21\times$ larger than \sse-128)} and slows scans to $387$\,ms.
\end{itemize}

\begin{table}[!t]
\centering\small
\caption{System cost at $N{=}1\mathrm{M}$ tuples: relation footprint, Q1 scan latency (median of $50$ queries), and habitat HNSW index.}\label{tab:cost}
\setlength{\tabcolsep}{2.2pt}
\resizebox{\columnwidth}{!}{%
\begin{tabular}{@{}lcccccc@{}}
\toprule
Physical Representation & Vecs & Data & Q1 Scan & Index & Build & Top-10 \\
 & /tup & (GB) & (ms) & d / GB & (s) & (ms) \\
\midrule
\textbf{\sse-128 (block-col.)} & \textbf{1} & \textbf{0.52} & \textbf{31} & \textbf{32 / 0.38} & 3.8 & \textbf{0.02} \\
\sse-768 (block-col.) & 1 & 3.08 & 32 & 32 / 0.38 & \textbf{3.4} & \textbf{0.02} \\
\sse-128 / 768 (row) & 1 & 0.52 / 3.08 & 61 / 136 & 32 / 0.38 & 3.8 & 0.02 \\
\midrule
B1 Labels (low acc.) & 1 & 3.08 & 19 & -- & -- & -- \\
\midrule
B5 Separate vectors & 6 & 3.68 & 24 & 32 / 0.38 & 3.8 & 0.02 \\
B6 CSN (masked 768d) & 1 & 3.09 & 280 & 768 / 3.33 & 17.7 & 0.06 \\
B6 CSN (active mat.) & 6 & 11.02 & 387 & 365 / 1.72 & 10.0 & 0.04 \\
\bottomrule
\end{tabular}}
\vspace{-2mm}
\end{table}

\stitle{Intrinsic Facet Dimensionality and Block Allocation}
Applying the validation-driven sizing rule of Section~\ref{sec:learning} ($\varepsilon{=}0.01$, $5$ training species held out for validation) across $d \in \{1,\dots,128\}$ reveals steep diminishing returns (Figure~\ref{fig:exp-sys}c): Diet correlation jumps from $-0.03$ ($d{=}1$) to $0.83$ at $d^*{=}3$ and plateaus completely through $d{=}64$, while Habitat Filter F1 rises from $0.28$ ($d{=}1$) to $0.68$ at $d^*{=}6$ and remains flat through $d{=}128$. Doubling-based exponential search requires nearly half as many training runs as dense grid search ($7$--$8$ vs.\ $11$--$14$) and selects neighboring sizes ($d^*{=}4$ and $8$) within $0.02$ of peak quality. This confirms that individual semantic facets possess low intrinsic dimensionality and leaves ample vector budget for schema evolution.

\stitle{Robustness Across Foundation Encoders}
Finally, replacing CLIP ViT-L/14 with EVA02-CLIP-L/14~\cite{sun2023evaclip} on AwA2, with zero changes to schema layout or hyperparameters, reproduces every conclusion: \sse blocks strongly capture Habitat, Diet, and Taxonomy ($\rho = 0.93$, $0.87$, $0.66$, vs.\ $0.29$, $0.28$, $0.23$ for plain EVA and $0.50$, $0.57$, $0.72$ for B1 Labels), composite query accuracy reaches $0.76$ Filter F1 and $0.58$ NDCG@10 (vs.\ $0.44$ / $0.36$ for labels and $0.53$ / $0.38$ for plain EVA), graded in-between similarity reaches $0.92$ (vs.\ $0.47$ for labels), and unseen prefix retrieval reaches $0.95$, $0.99$, and $0.99$ (vs.\ $0.79$ Order for MRL). \sses thus provide a robust, encoder-agnostic abstraction for relational-vector query processing.

\section{Illustrative Queries}\label{sec:queries}

The following queries illustrate the composition of these operators. We use animal and habitat records with the relevant \sse blocks and assume that blocks compared across relations are aligned. The taxonomy prefixes $E_{\mathrm{tax}}^{(1)}$, $E_{\mathrm{tax}}^{(2)}$, and
$E_{\mathrm{tax}}^{(3)}$ represent Order, Family, and Genus, respectively.
The SQL-like notation
$u\approx_\tau v$ abbreviates
$\simfn(u,v)\geq\tau$.
For grouping, {\tt GROUP BY}$_k$ specifies similarity-based clustering into $k$ groups, and {\tt REP(g,h)} represents group {\tt g} using up to $h$ representative tuples and their corresponding subgroup sizes, as described above.

\stitle{Q1: Independent filtering and ranking references}
Retrieve animals with habitats similar to that of a swan,
ranked by family-level similarity to a panda:
\insql{
    SELECT * FROM Animals AS t
    WHERE $E_{\mathrm{habitat}}(t)
           \approx_{\tau_h}
           E_{\mathrm{habitat}}(t_{\mathrm{swan}})$
    ORDER BY
    $\simfn(E_{\mathrm{tax}}^{(2)}(t),
                       E_{\mathrm{tax}}^{(2)}(t_{\mathrm{panda}}))$
    LIMIT 10.
}
The predicate and ranking use different blocks and different
reference objects. Varying $\tau_h$ changes the feasible
habitat neighborhood while preserving the ranking criterion.

\stitle{Q2: Composing ranges across facets}
Combine an Order-level predicate, a habitat similarity
interval, and a conventional attribute condition, then rank
by morphology:
\insql{
    SELECT * FROM Animals AS t
    WHERE $E_{\mathrm{tax}}^{(1)}(t)
           \approx_{\tau_t} c_{\mathrm{Carnivora}}$
    AND $\bot_h
         \leq \simfn(E_{\mathrm{habitat}}(t),
                                E_{\mathrm{habitat}}(t_{\mathrm{swan}}))
         \leq \top_h$
    AND t.ObservationYear >= 2020
    ORDER BY
    $\simfn(E_{\mathrm{morph}}(t),q_{\mathrm{morph}})$
    LIMIT 10.
}
The habitat interval can select a band of similarity rather
than all values above a lower bound. Each semantic condition
uses its own reference and thresholds, independently of
the morphology-based ranking.

\stitle{Q3: Joining through a shared semantic join key}
Find animal--park-site pairs whose habitat representations satisfy a selected similarity threshold:
\insql{
    SELECT A.id, P.id
    FROM Animals AS A VecJOIN ParkSites AS P
    ON $E_{\mathrm{habitat}}(A)
        \approx_{\tau_j}
        E_{\mathrm{habitat}}(P)$.
}
Each animal supplies the reference for a range over the ParkSites relation. Changing $\tau_j$ changes the required degree of similarity.

\stitle{Q4: Grouping without predefined categories.}
Group animals by habitat and represent each group using representative animal images:
\insql{
    SELECT REP(g,5), COUNT(*)
    FROM Animals AS t
    GROUP BY$_k$ $E_{\mathrm{habitat}}(t)$ AS g.
}
Each group is summarized by up to five representative images selected using the full embeddings, along with the number of tuples each image represents.

\stitle{Q5: Aggregation over semantic facets}
Measure the average and maximum habitat similarity of animals in the order Carnivora to the habitat of a swan, where $q_h=E_{\mathrm{habitat}}(t_{\mathrm{swan}})$:
\insql{
    SELECT AVG($\simfn(E_{\mathrm{habitat}}(t),q_h)$),
           MAX($\simfn(E_{\mathrm{habitat}}(t),q_h)$)
    FROM Animals AS t
    WHERE $E_{\mathrm{tax}}^{(1)}(t)
           \approx_{\tau_t} c_{\mathrm{Carnivora}}$.
}
Both the predicate and the aggregates are evaluated directly over SSE blocks, without requiring traditional attributes.

\stitle{Q6: Composing joins, predicates, grouping, and aggregation}
Identify animals with habitats compatible with sites in Sub-Saharan Africa, restrict them to the order Carnivora, group them at the Family level, and rank the groups by their average morphological similarity to a query object:

\sql{
    SELECT REP(g,5), COUNT(*),
           AVG($\simfn(E_{\mathrm{morph}}(t),q_{\mathrm{morph}})$) AS score
    \\FROM (
        SELECT DISTINCT A.*
        FROM Animals AS A VecJOIN ParkSites AS H
        ON $E_{\mathrm{habitat}}(A)
            \approx_{\tau_j} E_{\mathrm{habitat}}(H)$
        WHERE $E_{\mathrm{geo}}(H)
               \approx_{\tau_g} c_{\mathrm{SubSahAfrica}}$
        AND $E_{\mathrm{tax}}^{(1)}(A)
             \approx_{\tau_t} c_{\mathrm{Carnivora}}$
    ) AS t
    \\GROUP BY$_k$ $E_{\mathrm{tax}}^{(2)}(t)$ AS g
    \\ORDER BY score DESC
    LIMIT 10.
}

The habitat block determines join membership, geography and an Order-level prefix determine feasibility, a Family-level prefix determines grouping, and morphology determines the aggregate used for ranking the groups.
The result represents each group through representative animal images and their corresponding counts.

\section{Future Work and Final Remarks}
\label{sec:conclusion}

In this paper, we introduced Schema-Structured Embeddings (\sses) and the Relational-Vector Database (\rvdb) data and query models, together with learning methods and experimental evidence for their use. Developing a full \rvdb system raises several further research problems.
We briefly mention some of these problems below.

\stitle{Index Construction}
Range queries and approximate nearest neighbor (ANN) search are the core operations in \rvdbs. Various indices have been developed for these problems, including space-partitioning methods, tree-based data structures, and graph-based indices~\cite{pan2024survey}. Efficiently adapting these indices to constrained similarity search is already a challenging problem in conventional vector databases~\cite{li2025attribute}. In \rvdbs, the problem becomes more challenging since various database operations are modeled as similarity queries across different possibly independent subspaces. Moreover, the query point in the ranking space may correspond to an object that does not satisfy any of the filtering predicates. Designing indices that exploit the block and hierarchical structure of \sses to support efficient search under such geometric constraints is an important algorithmic research direction.

\stitle{Semantic Join Processing}
Vector similarity joins have been studied in prior work, including recent efforts such as DiskJoin~\cite{chen2025diskjoin} that enable large-scale similarity joins on disk-resident vectors.
Extending these techniques to \rvdbs is an interesting research direction.
In particular, join keys in \rvdbs may be specified over different blocks or hierarchical prefixes, and a join may involve several semantic predicates with different similarity thresholds.
Unlike ANN search, a semantic join may require reporting all matching pairs, as missing pairs may affect subsequent database operations such as grouping and aggregation.
Developing scalable join algorithms with explicit guarantees on the completeness of the join results is therefore of particular interest.

\stitle{Parameter Tuning and Query Optimization}
Queries in \rvdbs introduce several parameters, including similarity thresholds for predicates and joins, and the number of groups for semantic grouping. Specifying appropriate values for these parameters may not be straightforward for users. For example, the same similarity threshold may yield substantially different join cardinalities across facets or reference tuples. Automatically tuning these parameters based on user-specified requirements, such as result cardinality or semantic accuracy, is an interesting research problem. Moreover, once the query parameters are fixed, estimating the selectivity of geometric predicates and the cost of alternative execution plans is central to query optimization in \rvdbs.

\stitle{Query Accuracy}
Unlike traditional attributes, the attribute values in \sses are learned and may not perfectly reflect the intended semantics. Moreover, approximate query processing introduces another source of error. These errors may propagate through a sequence of database operations. Quantifying and controlling the accuracy of complex queries over \sses is an interesting direction for future work.

In this paper, we provided a common basis for addressing research challenges that enable relational-vector databases at scale. We hope this work supports further development of the algorithmic and systems foundations of \rvdbs, where learned semantic structure informs both data modeling and query processing.

\begin{acks}
This project was supported in part by the NSF grant \href{https://www.nsf.gov/awardsearch/show-award/?AWD_ID=2348919}{2348919}, CloudBank computation (\href{https://www.xras.org/public/requests/263909-ACCESS-CIS251215}{NSF ACCESS allocation CIS251215}), and an Amazon Research Award.
Any opinions, findings, and conclusions or recommendations expressed in this material are those of the author(s) and do not reflect the views of the supporting organizations.
\end{acks}
\balance
\bibliographystyle{ACM-Reference-Format}
\bibliography{ref,ref-proposal}

\end{document}